\documentclass[pra,
a4paper,
showkeys,
twocolumn,
superscriptaddress]{revtex4}
\usepackage{amssymb}
\usepackage{amsmath}
\usepackage{amsfonts}
\usepackage{graphicx}
\usepackage{bm}
\usepackage{color}
\usepackage{multirow}
\usepackage{natbib}
\usepackage{hyperref}
\usepackage[normalem]{ulem}
\usepackage{verbatim}
\def \be {\begin{equation}}
\def \ee {\end{equation}}
\def \bea {\begin{align}}
\def \eea {\end{align}}
\def \p {\partial}
\def \BEA {\begin{eqnarray}}
\def \EEA {\end{eqnarray}}

\def \BC {\begin{cases}}
\def \EC {\end{cases}}

\begin{document}

\title{
Stress-controlled Poisson ratio of a crystalline membrane: Application to graphene
}

\author{I.\,S.~Burmistrov}
\affiliation{L. D.~Landau Institute for Theoretical Physics, Kosygina
  street 2, 119334 Moscow, Russia}
\affiliation{Laboratory for Condensed Matter Physics , National Research University Higher School of Economics, 101000 Moscow, Russia}

\author{I.\,V.~Gornyi}
\affiliation{Institut f\"ur Nanotechnologie,  Karlsruhe Institute of Technology,
76021 Karlsruhe, Germany}
\affiliation{A. F.~Ioffe Physico-Technical Institute,
194021 St.~Petersburg, Russia}
\affiliation{\mbox{Institut f\"ur Theorie der kondensierten Materie,  Karlsruhe Institute of
Technology, 76128 Karlsruhe, Germany}}
\affiliation{L. D.~Landau Institute for Theoretical Physics, Kosygina
  street 2, 119334 Moscow, Russia}

\author{V.\,Yu.~Kachorovskii}
\affiliation{A. F.~Ioffe Physico-Technical Institute,
194021 St.~Petersburg, Russia}
\affiliation{L. D.~Landau Institute for Theoretical Physics, Kosygina
  street 2, 119334 Moscow, Russia}
\affiliation{Institut f\"ur Nanotechnologie,  Karlsruhe Institute of Technology,
76021 Karlsruhe, Germany}
\affiliation{\mbox{Institut f\"ur Theorie der kondensierten Materie,  Karlsruhe Institute of
Technology, 76128 Karlsruhe, Germany}}

\author{M.\,I.~Katsnelson}
\affiliation{Radboud  University,  Institute  for  Molecules  and  Materials,  NL-6525AJ  Nijmegen, The  Netherlands}

\author{J.\,H.~Los}
 \affiliation{Radboud  University,  Institute  for  Molecules  and  Materials,  NL-6525AJ  Nijmegen, The  Netherlands}

\author{A.\,D.~Mirlin}
\affiliation{Institut f\"ur Nanotechnologie,  Karlsruhe Institute of Technology,
76021 Karlsruhe, Germany}
\affiliation{\mbox{Institut f\"ur Theorie der kondensierten Materie,  Karlsruhe Institute of
Technology, 76128 Karlsruhe, Germany}}
\affiliation{L. D.~Landau Institute for Theoretical Physics, Kosygina
  street 2, 119334 Moscow, Russia}
\affiliation{Petersburg Nuclear Physics Institute, 188300, St.Petersburg, Russia}

\date{\today}

\begin{abstract}
We demonstrate that a key elastic parameter of a suspended crystalline membrane---the Poisson
ratio (PR) $\nu$---is  a non-trivial function of the applied stress $\sigma$ and of the system size $L$,
i.e., $\nu=\nu_L(\sigma)$.   We consider a generic two-dimensional membrane embedded into space of dimensionality $2+d_c$.   (The physical situation corresponds to $d_c=1$.)
A particularly important application of our results is to free-standing graphene.
We find that at a very low stress, when the membrane exhibits linear response, the PR  $\nu_L(0)$ decreases with increasing system size $L$ and saturates for $ L\to \infty$ at a value
which depends on the boundary conditions and is essentially different from the value $\nu=-1/3$
previously predicted by the membrane theory within a self-consisted scaling analysis.
By increasing $\sigma$, one drives a sufficiently large membrane (with the length $L$ much larger
than the Ginzburg length) into a non-linear regime characterized by a universal value of PR
that depends solely on $d_c$, in close connection with the critical index $\eta$ controlling the renormalization of bending rigidity. This universal non-linear PR acquires its minimum value
$\nu_{\rm min}=-1$ in the limit $d_c\to \infty$, when $\eta \to 0$.
With the further increase of $\sigma$, the PR changes sign and finally saturates at a positive
non-universal value prescribed by the conventional elasticity theory. We also show that one should distinguish between the absolute and differential PR ($\nu$ and $\nu^{\rm diff}$, respectively).
While coinciding in the limits of very low and very high stress, they differ in general:
$\nu \neq \nu^{\rm diff}$. In particular, in the non-linear universal regime, $\nu^{\rm diff}$ takes a universal value which, similarly to the absolute PR, is a function solely of $d_c$ (or, equivalently,
of $\eta$) but is different from the universal value of $\nu$. In the limit of infinite dimensionality
of the embedding space, $d_c\to \infty$ (i.e., $\eta \to 0$), the universal value of $\nu^{\rm diff}$   tends to $-1/3$, at variance with the limiting value $-1$ of $\nu$. Finally, we briefly discuss generalization of these results to a disordered membrane.
\end{abstract}
\maketitle

\section{Introduction}
\label{sec:intro}

One of the key elastic parameters of any material is the Poisson ratio (PR)
\be
\nu=-\frac{\varepsilon_y }{\varepsilon_x},
\ee
which is the coefficient governing the magnitude of transverse deformations $\varepsilon_y$ upon longitudinal stretching $\varepsilon_x$. Conventional materials contract in lateral directions when stretched, so that $\nu$ is typically positive. However, some exotic, so-called auxetic \cite{evans}, materials have negative $\nu$.
Although some examples of such materials, like  a pyrite crystal, were known long time ago \cite{pyrite},
the active study of auxeticity  started only  at  the end of 80's, triggered by the
observation of  stretching-induced transverse expansion in polyurethane foam \cite{foam}.
Since then, a negative  PR---both in intrinsic materials and in the artificially engineered structures---was reported  in a great number of publications 
(for recent review see Ref.~\cite{auxetic}). The increased interest to auxetic systems  is due to their unusual mechanical properties \cite{auxetic},
such as  increased sound velocity, which is proportional to $(1+\nu)^{-1/2}$, and enhanced  strength.

The purpose of this paper is to explore the PR in
graphene, which is a famous two-dimensional (2D) material displaying  
unique electrical and optical phenomena
\cite{Geim,Geim1,Kim,geim07,graphene-review,review-DasSarma,review-Kotov,book-Katsnelson,review-Katsnelson, book-Wolf,book-Roche}.
It also shows unusual elastic properties. 
In particular, free-standing graphene is a remarkable example of a crystalline 2D membrane
with an extremely  high bending rigidity ${\varkappa\simeq 1}$~eV.
A distinct feature of  such a membrane is the
existence of specific type of  dynamical and static out-of-plane  modes, known as flexural phonons (FP) \cite{Nelson} and ripples \cite{graphene-review,book-Katsnelson, review-Katsnelson}, respectively.

While the PR of graphene-related structures has been a subject of numerous experimental and  theoretical works, the results are by far not complete and largely conflicting. The experimental activities have focussed on graphene grown on a substrate, with the results for the PR spreading in the range between 0.15 and 0.45 for various substrates (see \cite{Cao2014,Politano2015} and references therein). These results were apparently influenced by the substrates in an essential way, so that it is difficult to extract from them an information about the PR of a freely-standing graphene. A direct measurement of the PR of suspended graphene remains a challenging prospect for future experimental work.

Let us briefly outline the state of the art in the computational analysis of graphene's PR. Early simulations  \cite{Zakharchenko2009}  predicted  that the PR of pristine graphene (that we term  ``clean'' below  as opposed to disordered graphene  with impurities and defects)  is   positive at ordinary experimental conditions  but  appears to become negative with  the temperature $T$ increasing above  a  quite large  value  ($T \gtrsim 1700$ K). Later work \cite{katsnelson16} supported the conclusion of a positive PR and found its variation with the system size in the interval 
$0.15 \lesssim  \nu \lesssim 0.3$. This value is close to value  $\nu\approx 0.17$ found in numerical simulations  \cite{Li} which did not take into account out-of-plane FP modes. 

On the other hand, a number of recent computational studies obtained negative values of the PR for
graphene \cite{Grima2015,Qin2017,Wan2017,Jiang2016,Ulissi2016,Park2016} and graphene-based engineered structures \cite{Wu2015,Ho2016}, thus demonstrating that graphene does exhibit auxetic properties. In particular,  it was  found that disorder is highly favorable for auxeticity of the membrane. Specifically, it was reported that introduction of local vacancy defects \cite{Grima2015} or artificially designed  ripples  \cite{Qin2017}  into a graphene flake leads to negative PR.  In a related work, Ref.~\cite{Wan2017}, it was found that the PR is negative in the graphene oxide at sufficiently large degree of oxidation. Another recent numerical work \cite{Ulissi2016} studied the dependence of PR on the applied stress and came to the conclusion that, while the PR is positive in the limit of zero stress and at very large stresses, it is negative in the intermediate range of stress.

To summarize, the available numerical simulations yield a positive PR of graphene under normal conditions but show that the PR becomes negative at high temperatures \cite{Zakharchenko2009} or in the presence of sufficiently strong disorder \cite{Grima2015,Qin2017,Wan2017}. The emergence of an auxetic behavior (negative PR) is qualitatively consistent with expectations based on the membrane theory \cite{Doussal}. Two decades ago, it was found in the framework of this theory that, when the membrane size $L$ exceeds the so-called Ginzburg length $L_*$ [see Eq.~\eqref{ginz1} below], elastic properties become universal and show an anomalous power-law scaling with $L$ controlled by a critical index $\eta$.  Recent years witnessed a revival of interest in elastic properties of membranes in the context of graphene and related 2D materials. It was shown, in particular, that anomalous elasticity of graphene  leads to anomalous temperature scaling of electric resistivity, formation of large-scale ripples, non-linear Hooke's law, and a negative thermal expansion coefficient. These theoretical results are in a decent agreement with experimental findings. A more detailed discussion, with references to relevant  theoretical and experimental works, is presented in Sec.~\ref{s2}.

Within the membrane theory, the PR in the limit of zero stress ($\sigma\to 0$, linear-response regime) was addressed in the framework of the self-consisted screening approximation (SCSA)
and  predicted \cite{Doussal} to be scale-independent  and given by a universal negative
value, $$\nu=-1/3 $$  (see also a discussion in Ref.~\cite{nelson13} and a review 
\cite{SCSA-Review}).
This result was recently rederived in Ref.~\cite{nelson15}.
On the numerical side, an early work  \cite{Zhang96} that performed molecular-dynamics simulations of a membrane with periodic boundary conditions yielded a negative PR, $\nu \approx -0.15$, twice smaller than the analytical value from Ref.~\cite{Doussal}. Later simulations, where no boundary constraints were imposed, yielded  considerably larger negative values of PR:  $\nu \approx-0.32$ for  phantom crystalline  membranes \cite{Bowick97} and  $\nu \approx-0.37$ for self-avoiding  crystalline membranes  \cite{Bowick2001}. The authors argued that these results are in agreement with the analytical predictions $\nu=-1/3 $ of Ref. \cite{Doussal}.
While both Ref.~\cite{Zhang96}, on one side, and Refs.~\cite{Bowick97,Bowick2001}, on the other side, obtained a negative PR, a clear difference in numerical values calls for an explanation (see also Ref.~\cite{footnote-nelson}).
If one believes in general applicability of the result $\nu=-1/3$, why did it fail in the case of Ref.~\cite{Zhang96}? And, if it fails there, under what conditions should it be applicable at all?

The situation becomes even more puzzling if one recalls positive values of PR obtained in numerical simulations for pristine graphene (at room temperature and for lowest values of stress), which should be contrasted to negative values of PR obtained in the earlier membrane simulations. A possible explanation is that the system size in graphene simulations was not sufficiently large. Indeed, it has been recently shown \cite{nelson15} that, with lowering system size, the PR (at $\sigma\to 0$) evolves towards a non-universal positive value following from the conventional theory of  elasticity. This crossover takes place at system sizes of the order of the Ginzburg length $L_*$. The value of $L_*$ in graphene at room temperature is
$40 \div 70$\AA, so that the condition $L>L_*$ appears to be usually satisfied in simulations. Thus, the conclusion that system sizes were not large enough seems somewhat surprizing.
Did some numerical factors intervene, thus shifting a crossover towards values of $L$ a few times larger than expected? And, finally, why did numerical simulations for disordered graphene show much more pronounced auxetic properties than for clean graphene?

In this paper, we develop a theory of the PR of graphene exemplifying a generic 2D crystalline membrane.
Our work extends previous studies in several essential directions.
First, we explore the dependence of PR of a finite-size membrane on the applied stress $\sigma$.
Second, we analyze the difference  between the absolute and differential values of PR ($\nu$ and $\nu^{\rm diff}$, respectively). We demonstrate that both $\nu$ and  $\nu^{\rm diff}$  are non-trivial functions of the applied uniaxial stress $\sigma$ and the system size $L$:
$\nu=\nu_L(\sigma), \nu^{\rm diff}_L(\sigma)$. While coinciding in the limits of very low and very high stresses,
$$\nu_L(0) =\nu_L^{\rm diff}(0)\quad \text{and}\quad \nu_L(\infty) =\nu_L^{\rm diff}(\infty),$$
in general, they differ, $\nu \neq \nu^{\rm diff}$.

\begin{figure}[h!]
\centerline{\includegraphics[width=0.8\linewidth]{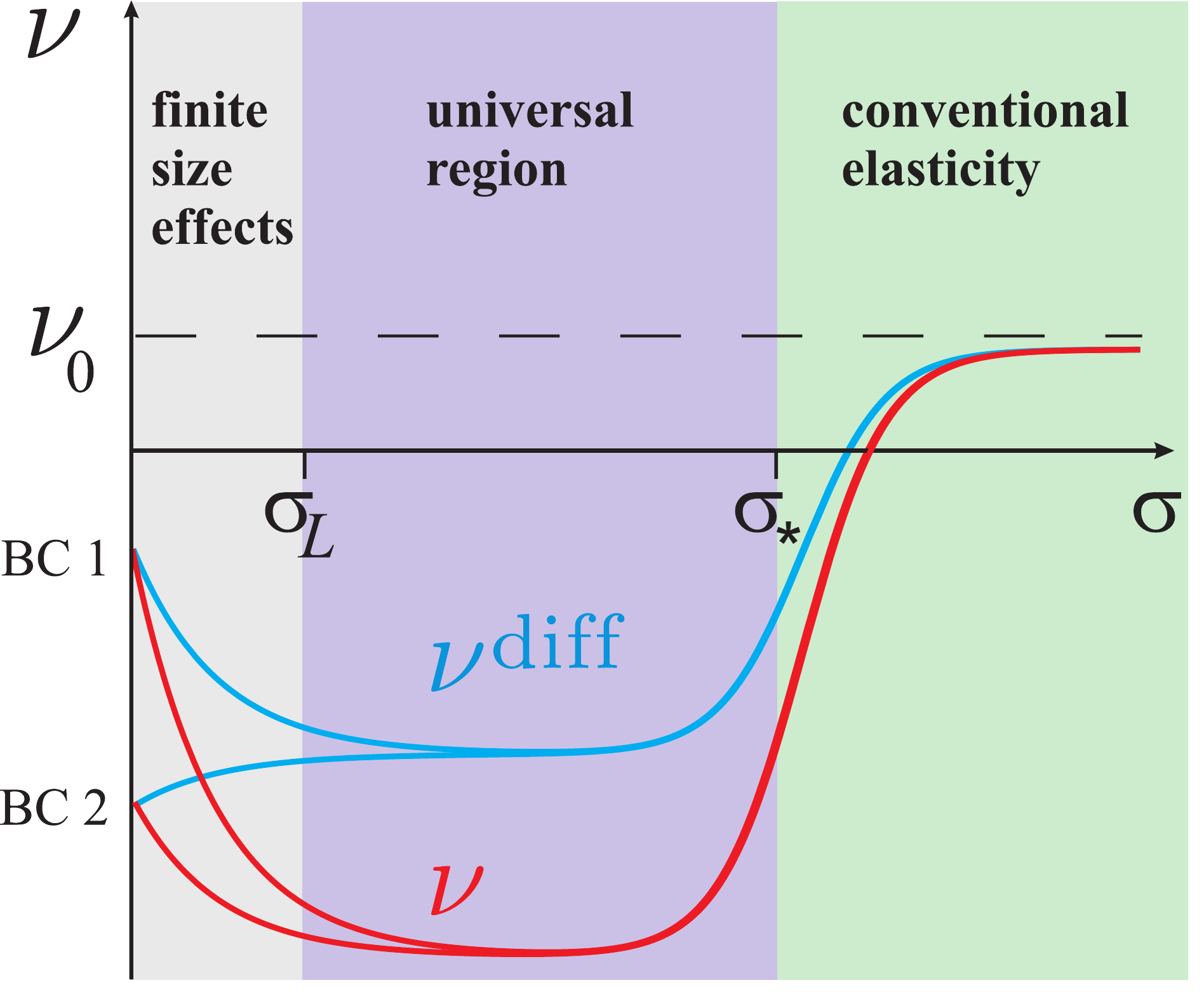}}
\caption{Schematic dependence of the absolute,  $\nu=\nu_L(\sigma)$, and differential,
$\nu^{\rm diff}=\nu_L^{\rm diff}(\sigma)$, PR in a crystalline membrane.
The characteristic scales of the stress, $\sigma_*$ and $\sigma_L$, are given by Eqs.~\eqref{sigma-star}  and \eqref{sigma-L}, respectively.  The values  of $\nu$ and $\nu^{\rm diff}$ coincide in the limits of very low and very high stresses,
$\nu_L(0) =\nu_L^{\rm diff}(0)$ and   $\nu_L(\infty) =\nu_L^{\rm diff}(\infty).$
At $\sigma \ll \sigma_L$, the PR is negative and depends on boundary conditions as indicated by curves BC~1 and BC~2 corresponding to the different boundary conditions.
At $\sigma  \gg \sigma_*$, the PR is positive and is given by the (material-dependent) value $\nu_0$ prescribed by the conventional elasticity theory. In the universal non-linear regime,
$\sigma_L \ll \sigma \ll\sigma_*$, the absolute and differential PR have different negative universal  values $\nu_{\infty}$ and $\nu_{\infty}^{\rm diff}$ which depend solely on the dimensionality $2+d_c$ of the embedding space, i.e., on the  critical index $\eta$. For $d_c \to \infty$ (i.e, $\eta \to 0$) these universal values exhibit the limiting behavior $\nu_{\infty}\to -1$ and  $\nu_{\infty}^{\rm diff} \to -1/3$, respectively.}
\label{Fig0}
\end{figure}

We will demonstrate that, for fixed finite $L$, the PR (both absolute and differential)
exhibits, with increasing $\sigma$, three distinct regimes (see Fig.~\ref{Fig0}).
In the limit  $\sigma \to 0$,  the absolute and differential PR coincide and depend on the system size.
The ``universality'' of PR in this regime has a very restricted meaning, even in the limit of large system size  ($L \gg L_*$), in contrast to the previous works \cite{Doussal,nelson15} that predicted a truly universal value  $-1/3$ of the PR. Specifically, while the linear-response PR of a large membrane is not sensitive to microscopic details of the system, it dramatically depends on the sample shape (aspect ratio) and on boundary conditions (BC), and can vary by an order of magnitude.

For $\sigma$ above a small, size-dependent value $\sigma_L \propto  1/L^{2-\eta}$, the system falls  into a {\it universal non-linear regime}  (provided that $L \gg L_*$) where the absolute and differential PR  are close to distinct universal values (the limit $L\to \infty$ is taken first):
$$\nu_\infty(\sigma \to 0) \neq \nu_\infty^{\rm diff}(\sigma \to 0).$$
The notion of universality here means independence from both microscopic details and BC.
On the other hand, these universal values of $\nu$ and $\nu^{\rm diff}$ do depend on the dimensionality $d_c$ and, in general, none of them is equal to $-1/3$, as discussed below.
With the further increase of  $\sigma$, the absolute and differential PR change sign and finally saturate  at the positive non-universal value $\nu_0$ prescribed by the conventional elasticity theory.

Importantly,  we show that the PR depends on the critical index $\eta$ (which is, in turn, a function of the dimensionality $d = 2+d_c$ of the embedding space) controlling the renormalization of bending rigidity and playing a key role in the crumpling and buckling transitions that can occur in a crystalline membrane (see discussion in Sec.~\ref{s2}). An analytical calculation of the PR for a physical 2D membrane in a three-dimensional space ($d_c=1$) thus encounters a severe obstacle: the absence of a small parameter that would control the analysis.  To overcome this difficulty, we consider the limit of large $d_c$, where  $\eta \simeq 2/d_c$ can be treated as a small parameter. We demonstrate that a small value of $\eta$ allows one to controllably calculate both the absolute and differential PR.
In particular, in the universal non-linear regime,  $\sigma_L  < \sigma < \sigma_*$
[for definitions of $\sigma_L$ and $\sigma_*$ see Eqs.~ \eqref{sigma-star} and \eqref{sigma-L} below],
we find
$$\nu \to  -1 \quad \text{and} \quad \nu^{\rm diff}\to -1/3$$
in the limit of large dimensionality $d_c\to \infty$  (i.e., $\eta\to 0$).
Leading corrections to these values are linear in $\eta$.
We also find analytically the values of  $\nu^{\rm diff}_L(0)=\nu_L(0)$ for various boundary conditions
in the $\eta\to 0$ limit.

For a physical membrane with $d_c \sim 1$, the value of $\eta$ is not at all small,
$\eta \approx 0.7 - 0.8$. Thus, the values of PR in all the above regimes will differ substantially from the corresponding values at $d_c \to \infty$ (or, equivalently, $\eta \to 0$). On the other hand, all the basic physical features of the  functions $\nu_L(\sigma)$ and $\nu^{\rm diff}(\sigma)$ are expected to be the same at $d_c = 1$ and at large $d_c$. Furthermore, we find a relatively small value of the numerical coefficient in front of the linear-in-$\eta$ term in the expansion for $\nu^{\rm diff}$. This suggests that the values of PR found in this work may serve as reasonable approximations for a physical membrane in a three-dimensional space.

We also discuss briefly the opposite limit, $\eta \to 1$,  which  is formally realized at
$d_c\to 0$ \cite{Doussal}. In this limit, effects of anomalous elasticity get suppressed with decreasing $d_c$, coming into play only at exponentially large scales, $L> \tilde{L}_*$, where
$\ln \tilde{L}_* \propto 1/d_c$.  (The definition  of  $ \tilde{L}_*$  is  given  in Sec.~\ref{eta-1})
For $L<\tilde{L}_*$, both the absolute and differential PR remain close to $\nu_0$.

Finally, we consider the PR of a disordered membrane. The physics is largely analogous in this case; however, the universality class is different. In particular, the index of anomalous elasticity has a distinct value, $\eta_{\rm dis} \simeq \eta/4$. As a result, the disordered membrane in the physical dimensionality ($d_c=1$) is much closer to the $d_c \to \infty$ (or, equivalently, $\eta \to 0$) limit than the clean one, which implies that disorder favors auxetic properties. We also show that in the linear-response ($\sigma \to 0$) regime the PR of a disordered membrane exhibits strong mesoscopic fluctuations.

\section{Anomalous elasticity of a generic membrane}
\label{s2}

We start with recalling basic notions of the anomalous elasticity of a generic crystalline membrane.
One of remarkable phenomena that may occur  in  such a  membrane is the crumpling transition (CT),
i.e., a transition between the flat and crumpled phases. The problem of crumpling has a close relation to the well known problem of thermodynamic stability of 2D crystals \cite{mermin,landau}
(see Refs.~\onlinecite{book-Katsnelson,review-Katsnelson} for a more recent discussion).

The underlying physics is the competition between thermal fluctuations and strong anharmonic coupling between in-plane vibration modes and FP \cite{Nelson}.
In contrast to the in-plane phonons with the linear dispersion,
the FP are very soft, $\omega_{ \mathbf{q}} \propto q^2$. Consequently,
the out-of-plane thermal fluctuations are unusually strong and tend to destroy the membrane by
driving it into the crumpled phase \cite{Nelson}.
The competing  effect is the anharmonicity
that  suppresses thermal fluctuations and, therefore,  plays here a
key role.
This question was intensively discussed more than two decades ago
\cite{Nelson,Nelson0,Crump1,NelsonCrumpling,david1,
buck,Aronovitz89, david2,
lower-cr-D2,
d-large,disorders,
disorder-imp,
Gompper91,RLD,
Doussal,
disorders-Morse-Grest,RLD1,
Bowick96}
in connection
with   biological membranes, polymerized layers, and inorganic surfaces.
The interest to this topic has been renewed more recently~\cite{Los-PRB-2009,eta1,Gazit1,Gazit2,Hasselmann,kats1,kats2,Amorim,kats3}
after discovery of graphene.
It was found \cite{Nelson0,Crump1,NelsonCrumpling,david1,Aronovitz89,buck,david2} that the anharmonic coupling of
in-plane and out-of-plane phonons stabilizes the membrane for not too high temperatures $T$.
This class of problems is under active investigation now since measurement of the elasticity of free-standing graphene is accessible to current experimental techniques \cite{lee, metten,blees15,lopez-polin15,nicholl15}. An additional interest to this topic is due to the significant effect of the FP and ripples  on electical and thermal conductivities of graphene (see Refs.~\cite{my-cond,Zhu} and references therein).

The CT temperature $T_{\rm cr}$ is  proportional to the bending rigidity \cite{my-crump,my-hooke} and, consequently, is  very high for graphene (of the order of several eV). Because of the high
value of $\varkappa$, {\it clean} graphene remains flat up to all realistic  temperatures,
$T\lesssim \varkappa$, and the CT can not be directly observed.
Remarkably, a crystalline membrane is predicted to show a  critical behavior even very far from the CT  transition point, deep in the flat phase. This is connected with a
strong renormalization of the bending rigidity \cite{Aronovitz89,lower-cr-D2,Doussal},
$\varkappa\to\varkappa_q$, for sufficiently small wave vectors
$q\ll q_*$
according to the RG equation \cite{Aronovitz89,lower-cr-D2,Doussal},
\be
 {d\varkappa}/{d \Lambda}= \eta \varkappa \,\,\,\Rightarrow \,\,\,\varkappa_q=\varkappa_0 \left( {q_*}/{q}\right)^\eta.
\label{kappa0}
 \ee
Here $\Lambda= \ln(q_*/q)$, $\eta$ is the anomalous dimension of the bending rigidity
(critical index of CT), $q_*$ is the inverse Ginzburg length,
\be
q_* =\frac{1}{L_*}\simeq {\sqrt{d_c\,\tilde{\mu}\,T }}/{\varkappa_0},
\label{ginz1}
\ee
 ${\tilde{\mu}=3\mu_0(\mu_0+\lambda_0)/[8\pi(2\mu_0+\lambda_0)]}$
(see, e.g.,
Ref.~\cite{my-crump}),
$\mu_0, \lambda_0$ are the bare in-plane elastic constants (Lam{\'e} coefficients), and  $\varkappa_0$ the bare bending rigidity.
The critical exponent $\eta$ was determined within several
approximate analytical schemes \cite{david1,david2,Aronovitz89,Doussal,eta1}  none of them is controllable  in the physical case of a 2D membrane embedded in the three-dimensional space.
Numerical simulations for latter case yield
$\eta=0.60 \pm 0.10$ \cite{Gompper91},
$\eta=0.72 \pm 0.04$ \cite{Bowick96}      and $\eta=0.85$   \cite{Los-PRB-2009}.

As a consequence of strong anharmonicity,  key characteristics of graphene, such as the conductivity at the Dirac point \cite{my-cond} and elastic moduli \cite{my-crump},  show a non-trivial power-law scaling  with the system size and temperature. Physically, this scaling manifests the tendency of the membrane     to the flat phase with increasing system size for temperatures below the crumpling transition temperature. [The latter condition is always satisfied for a graphene membrane, see Eq.~\eqref{Tcr} below.]
One of the most important consequences is that the linear Hooke's law fails even
in the limit of an  infinitesimally small tension \cite{buck, david2,lower-cr-D2, nicholl15, nelson15, katsnelson16, my-hooke}.
Specifically, the deformation  $\Delta L$ of a membrane subjected to a small stretching tension
$\sigma>0$ scales as ${\Delta L\propto\sigma^\alpha}$,
with a non-trivial exponent $\alpha$  which is expressed in terms of the critical index $\eta$ as
\be
\alpha = \frac{\eta}{2-\eta}.
\label{alpha}
\ee
In the opposite case, $\sigma<0$, $\Delta L < 0$, the membrane undergoes a buckling transition \cite{buck}, with $\alpha$ being the critical index of this transition.  Another remarkable  manifestation of the anomalous elasticity characterized by the critical scaling in the flat phase is that
the thermal expansion coefficient of graphene is negative and depends on $\eta$ \cite{my-quantum}.

Influence of static disorder on membrane elasticity has been discussed since early works \cite{disorders,disorder-imp,disorders-Morse-Grest}. In recent years, this question has attracted a great deal of attention in connection with ripples---{\it static} out-of-plane deformations induced by  disorder. In particular, in a recent paper by three of the authors~\cite{my-crump},
a theory of rippling and crumpling in {\it disordered} free-standing graphene was developed.
The coupled RG equations describing the combined flow of the bending rigidity and disorder strength
were derived and rippling in the flat phase was explored. It was  shown that the static disorder
can strongly affect  elastic properties of the membrane. In particular, the corresponding scaling exponent turns out to be four times smaller than in the clean case, $\eta_{\rm dis} \simeq \eta/4$.
It was also demonstrated \cite{my-hooke} that, similarly to the clean case, the linear  Hooke's law in {\it disordered} graphene breaks down at low stresses. Importantly, both  in the clean and disordered cases,  $\alpha $ is expressed in a simple way via the critical index $\eta$  but the values of $\alpha$ for clean and  disordered graphene are different and given by $\alpha(\eta)$ and $\alpha(\eta_{\rm dis})$, respectively.

These findings imply that FP and ripples can be studied on equal footing.
For weak disorder, FP dominate, while at sufficiently strong disorder the anomalous elasticity of
graphene is fully determined by the static random ripples. The non-linearity of
elasticity of graphene found in Ref.~\cite{my-hooke} is in agreement with recent experimental findings~\cite{lopez-polin15,nicholl15}. Related theoretical results have been recently obtained for
clean membranes in the ribbon geometry~\cite{nelson15}.

In this work, we mainly focus on the study of the PR of clean graphene. However, based on the similarity of the problems of FP and ripples, we supplement this analysis by a discussion of the disordered case.

\section{Balance equations for membrane}
\label{s3}

In this Section, we extend the theory of non-linear elasticity of 2D membranes \cite{my-hooke}
to study the PR. For the sake of generality, we consider (following earlier studies of membranes)
a  more general case of a membrane with dimension $D=2$ embedded in the $d$-dimensional space with
$d = 2 + d_c >2$. The physical situation corresponds to $d=3$.

The starting point of our analysis is the energy functional
\BEA
E&=& \int d^2x \left[\frac{\varkappa_0 }{2}(\Delta{\mathbf r})^2
+ \frac{\mu_0}{4}(\partial_\alpha{\mathbf r}\partial_\beta{\mathbf r} - \delta_{\alpha\beta})^2
  \right.
\nonumber
\\
&+&
\left.\frac{\lambda_0}{8}(\partial_\gamma{\mathbf r}\partial_\gamma{\mathbf r} - D)^2 \right],
\label{E}
\EEA
where  $\varkappa$ is the bare bending rigidity, while $\mu$ and $\lambda$ are
the in-plane coupling constants. The $d$-dimensional vector
$\mathbf r= \mathbf r(\mathbf x)$ describes
a point on the membrane surface and depends on the 2D coordinate
$\mathbf x=x \mathbf e_x +y \mathbf e_y$ that parametrizes the
membrane. Here, $\mathbf e_x $ and $\mathbf e_y$ are the unit vectors in the reference plane.
The vector $\mathbf r$ can be split into
\be \mathbf r=
\xi_{ij}x_i  \mathbf e_j + \mathbf u +\mathbf h,  \label{param} \ee
where vectors
$\mathbf u=(u_x,u_y),~ \mathbf h =(h_1,...,h_{d_c}) $
represent in-plane and out-of-plane displacements, respectively.
Homogeneous stretching of membrane in $x$ and $y$ directions is described by the tensor $\xi_{ij}$.
For isotropic deformations $\xi_{ij}=\xi \delta_{ij}$. In the absence of external tension, within
the mean-field approximation, the stretching factor $\xi$ equals to unity. Fluctuations (in particular, out-of-plane deformations) lead to a decrease of $\xi$, so that at finite $T$ the stretching factor becomes smaller than unity.

Here, we consider the reaction of the membrane to external forces applied in $x$ and $y$ directions.
For simplicity, we do not discuss shear deformations.
We thus assume that $\xi_{ij}$ has two non-zero spatially-independent components:
\be
\xi_{xx}=\xi_x,  \quad \xi_{yy}=\xi_y.
\label{deformations}
\ee
Details of the derivation of the free energy $F$  are relegated to
Appendix \ref{App-balance}, where
we obtain Eq.~\eqref{F} for  $F$ as a function of global deformations $\xi_x$ and $\xi_y$.
One can find the balance equation by differentiating $F$ with respect to deformations,
$\sigma_\alpha=L^{-2} \p F/\p \xi_\alpha$, where $\sigma_\alpha$ are components of the external stress applied to the membrane. As a result, we get
\be
\sigma_\alpha =
 M_{\alpha\beta} \frac{\xi_\beta^2-1+K_\beta}{2},
\label{tensor-hooke}
\ee
where
\be
K_\alpha=  \langle  K_\alpha^0 \rangle =\int \frac{ d\mathbf x}{L^2} \langle (\p_\alpha \mathbf h)^2\rangle  ,
\quad \alpha=(x,y)
\label{K-ave}
\ee
are the bulk-averaged anomalous deformations $K_\alpha^0$ [see Eq.~\eqref{K-L}],
also  averaged over the Gibbs distribution with the energy functional Eq.~\eqref{E}
under the fixed value of the external tension. The
matrix of elastic constants reads
\be
\hat M =\left(
          \begin{array}{cc}
            2\mu_0+\lambda_0 & \lambda_0 \\
            \lambda_0 & 2\mu_0+\lambda_0 \\
          \end{array}
        \right).
\label{M}
\ee

Unusual anomalous properties of membranes which are not captured by the conventional elasticity
theory are connected with the shrinking of the effective area of the membrane (projected area) caused by
the transverse fluctuations \cite{lifshitz}.
The effect of transverse fluctuations is described by the anomalous deformations $K_\alpha$
which are (by definition) some  functions of global deformations $\xi_x$ and $\xi_y$, and, consequently [via Eqs.~\eqref{tensor-hooke}], of the stress:  $K_\alpha =K_\alpha (\sigma_x,\sigma_y)$.
The anomalous deformations can be expressed in terms of the correlation function
$G_\mathbf q$  of FP:
\be
K_\alpha=  d_c\int \frac{d^2\mathbf q}{(2\pi)^2} ~q^2_\alpha G_\mathbf q,
\label{Kk}
\ee
where $G_\mathbf q$ is defined as
\be
\langle h_{\alpha,\mathbf q}   h_{\beta,-\mathbf q} \rangle =\delta_{\alpha\beta} G_\mathbf q.
\ee
Within the harmonic approximation, the bending rigidity is given by its bare value and the correlation function reads:
\be
 G_\mathbf q^{\rm har}  = \frac{T}{ \varkappa_0 q^4 + \sum_\alpha \sigma_\alpha q^2_\alpha}.
 \label{harmonic}
 \ee
The term  $\sum_\alpha \sigma_\alpha q^2_\alpha$ in the denominator accounts for a finite stress applied to the membrane.
The anharmonic coupling between FP and in-plane modes leads to essential modification of the correlation
function \eqref{harmonic}.
In particular, the account of this coupling within the random phase approximation (RPA)
scheme leads to a replacement of the bare value $\varkappa_0$ with the renormalized scale-dependent bending rigidity, $\varkappa_q$, in the denominator of Eq.~\eqref{harmonic} [see Eq.~\eqref{G-SCSA-1} below].

Equations \eqref{tensor-hooke}  are the basis for our further study.
For the sake of generality, we also present in Appendix \ref{App-balance} the balance equations and expressions for elastic moduli for a membrane of dimensionality $D\neq 2$.

The dependence of the anomalous deformations $K_\alpha$ on the applied stress is the key point for  further consideration.
Let us split $K_\alpha$ in two parts,
\be
K_\alpha(\sigma_x,\sigma_y)=K(0)-\delta K_\alpha(\sigma_x,\sigma_y).
\ee
where
\be K(0)=K_x(0,0)=K_y(0,0)
\ee
is the anomalous deformation at zero stress.
Physically, the deformation $K(0)$ arises because of the shrinking of the membrane
in the longitudinal direction caused by temperature-induced transverse fluctuations.
In a clean membrane, this deformation is proportional to the temperature \cite{david2,my-crump,my-hooke}, $K(0) =T/T_{\rm cr}$, and
can be fully incorporated in the renormalization of $\xi_\alpha$ in the unstressed membrane:
\be
\xi_\alpha^2 -1 \ \longrightarrow \ \xi_\alpha^2 - \xi_0^2,
\ee
where
\be
 \xi_0^2=1- K(0)=1-\frac{T}{T_{\rm cr}}
 \ee
and $T_{\rm cr}$ is the critical temperature for the crumpling.

We will assume that the membrane is in the flat phase far from the CT,
\be T\ll T_{\rm cr}\propto \varkappa.
\label{Tcr}
\ee
For graphene, where $\varkappa \approx 1~$eV, this is the case at all realistic temperatures.
Then, $$\xi_0\approx 1$$ and $$\xi_\alpha^2-\xi_0^2\approx 2\varepsilon_\alpha,$$ where $$\varepsilon_\alpha =\xi_\alpha-\xi_0$$ is a small deformation. Equation \eqref{tensor-hooke} then yields
 \BEA
 \varepsilon_x&=&\frac{\sigma_x-\nu_0\sigma_y}{Y_0} +\frac{\delta K_x(\sigma_x,\sigma_y)}{2},
 \label{eps-x}
 \\
 \varepsilon_y&=&\frac{\sigma_y-\nu_0\sigma_x}{Y_0} +\frac{\delta K_y(\sigma_x,\sigma_y)}{2},
 \label{eps-y}
 \EEA
 where
 \be
 Y_0=\frac{4\mu_0(\mu_0+\lambda_0)}{2\mu_0+\lambda_0},\quad \nu_0=\frac{\lambda_0}{2\mu_0+\lambda_0}
\label{bare}
 \ee
are the bare values of the Young modulus and of the PR, respectively.

Equations \eqref{eps-x} and \eqref{eps-y}  represent the general balance equations for a crystalline membrane (in the absence of  shear deformations) deep in the flat phase.  The key new ingredients of these equations, compared to balance equations of the conventional elasticity theory, are anomalous deformations   $\delta K_\alpha$. Physically, the anomalous deformations at finite $\sigma$ account for uncrumpling, i.e., ``ironing'' of the membrane by the external stress. We will start with a phenomenological approach to anomalous elasticity by considering $\delta K_\alpha(\sigma_x,\sigma_y)$ to be a given function of
$\sigma_x$ and $\sigma_y$. The analytical expressions for these deformations will be presented later [see Eq.~\eqref{deltaq}].

We proceed now by briefly reminding the reader on implications of the anomalous elasticity in the case of an isotropic stress and  then by giving a general definition of  the absolute and  differential  PR. We will assume here  the limit of a large system size, $L\to \infty$, taken at a given value of stress. Finite-size effects will be analyzed in Sec.~\ref{s5}.

\subsection{Isotropic deformation}

Of central importance for anomalous elasticity is a strong renormalization of the elastic constants
by anomalous deformations $\delta K_\alpha$. To explain this point, we first note that for isotropic deformations $\sigma_x=\sigma_y=\sigma$, $\varepsilon_x=\varepsilon_y=\varepsilon$, $\delta K_x=\delta K_y=\delta K(\sigma)$. The balance equations then reduce to the following equation relating
$\sigma$ and $\varepsilon$:
\be
\varepsilon= \frac{\sigma}{k_0}+\frac{\delta K(\sigma)}{2},
\ee
in agreement with Refs.~\cite{david2,my-hooke}.
Here, $k_0=2(\mu_0+\lambda_0) \sim \mu_0$ is the bare in-plane stiffness.
(Here and below, we assume in all order-of-magnitude estimates that the bare elastic constants have the same order of magnitude: $\lambda_0 \sim \mu_0 \sim Y_0 \sim k_0$).
As shown in Ref.~\cite{my-hooke}, the renormalized stiffness
$k_{\rm eff}=\p \sigma/\p\varepsilon$ coincides with $k_0$ for large $\sigma$ but
is suppressed  in a power-law way,
 \be
 k_{\rm eff} \sim k_0 \left(\frac{\sigma}{\sigma_*}\right)^{1-\alpha},
 \label{Hooke}
 \ee
 for $\sigma <\sigma_*,$ where
\be
 \sigma_* \simeq \varkappa_0 q_*^2 \simeq \frac{ \mu_0 T}{\varkappa},
 \label{sigma-star}
\ee
and $\alpha$ is a critical index of buckling transition, which can be expressed in terms of  $\eta$ according to Eq.~(\ref{alpha}).
For $\sigma \ll \sigma_*$, the deformation $\epsilon$ is fully determined by the anomalous contribution:
$\varepsilon \simeq \delta K(\sigma)/2$.
The anomalous Hooke's law (\ref{Hooke}) originates form the critical scaling of the bending rigidity,
Eq.~(\ref{kappa0}).

\subsection{Absolute Poisson ratio}

Let us now consider a membrane subjected to an  uniaxial stress  in $x$-direction:
\be
\sigma_x=\sigma, \quad \sigma_y=0.
 \label{uni}
 \ee
The balance equations become
 \BEA
 \varepsilon_x&=&\frac{\sigma}{Y_0} +\frac{\delta K_x(\sigma,0)}{2},
 \\
 \varepsilon_y&=&-\frac{\nu_0\sigma}{Y_0} +\frac{\delta K_y(\sigma,0)}{2}.
 \EEA
Resolving these equations, we find the Young modulus and the absolute PR:
\BEA
&&Y=\frac {\sigma}{\varepsilon_x}=\frac{Y_0}{1+Y_0 \delta K_x /2\sigma},
\label{Y-absolute}
\\
&&\nu=-\frac{\varepsilon_y}{\varepsilon_x}=\frac{\nu_0- Y_0 \delta K_y/2\sigma }{1 + Y_0 \delta K_x/2\sigma}.
\label{nu-absolute}
\EEA
(Here and below  we omit arguments of $\delta K_\alpha$ for the sake of compactness.)
One can easily check  that the Young modulus and the absolute PR are connected by conventional expressions [see Eqs.~  \eqref{bare}] with the effective Lam{\'e} coefficients $\lambda$ and $\mu$ found from the equations
\BEA
&& \mu=\frac{\mu_0}{1+\mu_0 \delta K_-/\sigma},
\label{muK}
\\
&& \mu+\lambda=\frac{\mu_0+\lambda_0}{1+2(\mu_0+\lambda_0) \delta K_+/\sigma},
\label{mulK}
\EEA
where
\be
\delta K_+=\frac{\delta K_x+\delta K_y}{2},\quad \delta K_-=\delta K_x-\delta K_y.
\label{dKpm}
\ee
In order to clarify the physical meaning of $\delta K_\pm$, we notice that the
matrix $M_{\alpha \beta}$  defined by Eq.~(\ref{M}) is diagonalized by the transformation from $\sigma_x,\sigma_y$ to $\sigma_\pm= \sigma_x \pm \sigma_y$.
Physically, this means that  within the conventional elasticity  there are two types of deformations:
(i) isotropic deformations  with  $ \epsilon_x =\epsilon_y$ and (ii) deformations with  $ \epsilon_x =-\epsilon_y$, which correspond to eigenvalues $2(\mu_0+\lambda_0)$ and $2 \mu_0$ of  the matrix $\hat M$, respectively. Equations~\eqref{muK} and \eqref{mulK} show how  these eigenvalues are modified by the anomalous deformations.

The absolute PR is expressed in terms of $\delta K_{\pm}$ as follows:
  \be
  \frac{\nu+1/3}{Y}= \frac{\nu_0+1/3}{Y_0} +\frac{\delta K_-  - \delta K_+}{3\sigma}.
  \label{absPR-dK}
  \ee
In the limit of large anomalous deformations,
$ \delta K_{\alpha}/\sigma  \gg 1/Y_0$, we find $Y  \simeq 2\sigma/\delta K_{x}$ and
\BEA
&&\lambda=-\frac{\mu}{2} \left( 1+ \frac{\delta K_+-\delta K_-}{\delta K_+}\right), \label{inv-manifold}
\\
&&\nu = -\frac{\delta K_{y}}{\delta K_{x}}=-\frac{1}{3} + \frac{4}{3}~\frac{ \delta K_- - \delta K_+}{2\delta K_{+} +\delta K_{-}}.
\label{nueff-abs}
\EEA
We see that the Lam{\'e} coefficients belong to so-called invariant manifold \cite{Doussal},
$\lambda=-\mu/2$, and the absolute PR equals to $-1/3$ only provided that $\delta K_- = \delta K_+$.
However, as we demonstrate below, the latter equation is not satisfied even in the
limit $d_c \to \infty$.

\subsection{Differential Poisson ratio}

Next we consider the response of a membrane with respect to small variations $\delta \sigma_x$
and $\delta \sigma_y$.  Substituting  $\sigma_x=\sigma_x^0+\delta \sigma_x$ and  $\sigma_y=\sigma_y^0+\delta \sigma_y$ into  Eqs.~\eqref{eps-x} and \eqref{eps-y},  we find the linear-in-$\delta\sigma_\alpha$ variations of deformations
\BEA
 \delta \varepsilon_x&=&\frac{\delta \sigma_x-\nu^{\rm diff} \delta \sigma_y}{Y^{\rm diff}},
 \label{dif-eps-x}
 \\
 \delta \varepsilon_y&=&\frac{\delta \sigma_y-\nu^{\rm diff} \delta \sigma_x}{Y^{\rm diff}}.
 \label{dif-eps-y}
 \EEA
 Here
 \BEA
 && Y^{\rm diff}=\frac{Y_0}{1+Y_0\Pi_{xx}/2T}, \label{Y-effective}
 \\
 &&\nu^{\rm diff}=\frac{\nu_0 -Y_0 \Pi_{xy}/2T}{1+Y_0 \Pi_{xx}/2T},\label{nu-effective}
 \EEA
are the ``differential'' values of the Young modulus and the PR, respectively,
and
 \be
 \Pi_{\alpha\beta}=- T\frac{\p  K_{\beta} (\sigma_x^0,\sigma_y^0)}{\p \sigma_\alpha^0}= T\frac{\p  \delta K_{\beta} (\sigma_x^0,\sigma_y^0)}{\p \sigma_\alpha^0}
 \label{pi alpha beta}
  \ee
are the zero-momentum components of the  polarization operator
$\Pi^{\mathbf q}_ {\alpha\beta\gamma\delta}$
(which is a rank-four tensor)  \cite{lower-cr-D2,Doussal,my-crump},
 \be
 \Pi_{xx}= \Pi_{xxxx}^{\mathbf{q} \to 0 },\quad  \Pi_{xy}= \Pi_{xxyy}^{\mathbf{q} \to 0 }.
 \ee

The values of the  differential Young modulus and PR are connected by conventional relations of the form \eqref{bare} with the screened values of the Lam{\'e} coefficients,
$\mu^{\rm diff}$ and $\lambda^{\rm diff}$,
which can be found from
\begin{eqnarray}
\mu^{\rm diff}&=&\frac{\mu_0}{1+\mu_0 \Pi_-/T},
\label{mupm}
\\
\mu^{\rm diff}+\lambda^{\rm diff}&=&\frac{\mu_0+\lambda_0}{1+2(\mu_0+\lambda_0) \Pi_+/T},
\end{eqnarray}
where
\be
\Pi_+=\frac{\Pi_{xx}+\Pi_{xy}}{2}, \quad \Pi_-=\Pi_{xx}-\Pi_{xy}.
\ee
The physical sense of $\Pi_\pm$ is analogous to that of $\delta K_{\pm}$
discussed below Eq.~(\ref{dKpm}).
Equations~\eqref{mupm} describe the ``differential screening'' of the two eigenvalues
of the matrix $\hat M$.

One can express the differential PR in terms of $\Pi_{\pm}$ in a way similar to Eq.~\eqref{absPR-dK}:
  \be
  \frac{\nu^{\rm diff}+1/3}{Y^{\rm diff}}= \frac{\nu_0+1/3}{Y_0} +\frac{\Pi_- - \Pi_+}{3 T}.
  \ee
In the limit of strong screening, $\Pi_{\alpha\beta} \gg 1/Y_0$,
we find $Y^{\rm diff} \simeq 2T/\Pi_{xx}$ and
\BEA
&&\lambda^{\rm diff}= - \frac{\mu_{\rm diff}}{2} \left(1+ \frac{\Pi_+- \Pi_-}{\Pi_+}\right)
\label{lam-diff}
\\
&&\nu^{\rm diff} = -\frac{\Pi_{xy}}{\Pi_{xx}}=-\frac{1}{3} + \frac{4}{3}~\frac{ \Pi_- - \Pi_+}{2\Pi_{+} +\Pi_{-}}.
\label{nueff}
\EEA
For $\Pi_- = \Pi_+$, the  Lam{\'e} coefficients belong to the invariant manifold,
$\lambda^{\rm diff} =-\mu^{\rm diff} /2$, and the PR equals to $-1/3$.
As we will show below, this happens only in the limit $d_c \to \infty$.

 The differential Young modulus and the differential PR, as well as the polarization tensor, are  functions of the initial stress $(\sigma_x^0,\sigma_y^0)$.
 In the rest of the paper, when discussing  the differential PR, we assume an isotropic
 case, $\sigma_x^0=\sigma_y^0=\sigma$, i.e.,
  $Y^{\rm diff}= Y^{\rm diff}(\sigma,\sigma),\  \nu^{\rm diff}=\nu^{\rm diff}(\sigma,\sigma)$ and $\Pi_{xx}=\Pi_{xx}(\sigma,\sigma), \ \Pi_{xy}=\Pi_{xy}(\sigma,\sigma)$.
It is worth noticing that  Eqs.~ \eqref{Y-effective} and \eqref{nu-effective} for the differential response in the case of an isotropic stress $\sigma$  can be obtained
from  Eqs.~\eqref{Y-absolute} and \eqref{nu-absolute} for the absolute response to
an uniaxial stress $\sigma$ by the replacement
\be
 \frac{\delta K_x(\sigma,0)}{\sigma} \to \frac{\Pi_{xx}(\sigma,\sigma)}{T},  \quad
\frac{\delta K_y(\sigma,0)}{\sigma}\to \frac{\Pi_{xy}(\sigma,\sigma)}{T}.
\label{pi1}
\ee

\section{Calculation of Poisson Ratio}

As a first step of calculation of the PR, one can integrate out the in-plane modes in the energy functional \eqref{E}, thus arriving (in the absence of the external tension) to a functional
\BEA
\frac{E[\mathbf h]}{T}\!&=&\!\frac{\varkappa}{2T}\int (dk) k^4 |\mathbf h_\mathbf k|^2
\label{F} \\
\nonumber
&+&\!\!\frac{1}{8}\!\int (dk dk' dq) R_\mathbf q(\mathbf k, \mathbf k')
\left( \mathbf h_{\mathbf k+\mathbf q} \mathbf  h_{-\mathbf k} \right)
\left( \mathbf h_{-\mathbf k'-\mathbf q} \mathbf  h_{\mathbf k'} \right),
\EEA
which depends on $\mathbf h$ fields only \cite{Doussal}.
Here, we use a short-hand  notation $(dk)=d^2\mathbf k/(2\pi)^2$.
The anharmonic interaction between $\mathbf h$ and $\mathbf u$ fields
is encoded in $E[\mathbf h]$ in the ${\mathbf h}^4$ interaction   term with the   coupling
\be
R_\mathbf q (\mathbf k,\mathbf k')=Y_0 \frac{[\mathbf k\times \mathbf q]^2}{q^2}\frac{[\mathbf k'\times \mathbf q]^2}{q^2}.
\label{R2D}
\ee
Hence, the bare Young modulus serves as a bare coupling constant.
The bare propagator (which is exact in the absence of interaction, $R=0$)
is given by Eq.~\eqref{harmonic}.

The interaction coupling constants get screened
in analogy with conventional charges in a media with a finite polarizability.
Within the RPA,
one replaces  $Y_0$ with
    \be
        Y_\mathbf q  = \frac{Y_0}{1+Y_0\Pi^\mathbf q_{xxxx}/T}
          \ee
in Eq.~\eqref{R2D}, so that the screened coupling constant is $\mathbf q$-dependent.
For large systems, $L \gg L_*$, the properties of the membrane are determined by the
infrared  universal region,
$q\ll 1/L_*$, where interaction is proportional to the inverse polarization operator \cite{Doussal}
(see also Ref.~\cite{my-crump}):
\be
Y_\mathbf q  = {T}/{\Pi^{\mathbf q}_{xxxx}}.
          \label{Nq}
          \ee
We note that the bare coupling $Y_0$ drops out from the expression (\ref{Nq})
for the interaction in this regime.

The next step is to study  the nonlinear $\mathbf{h}^4$ model with the screened
interaction \eqref{Nq}.
In the absence of the external tension, the correlation functions of $\mathbf h$-fields
scales as follows \cite{Nelson}:
\be
G_\mathbf{q}(\sigma=0)=\frac{T}{\varkappa_q  q^4}\propto \frac{1}{q^{4-\eta}}.
\label{G-SCSA}
\ee
Equation (\ref{G-SCSA}) differs from  Eq.~\eqref{harmonic} with $\sigma=0$ by a replacement of the bare value of the  bending rigidity $\varkappa_0$ with the value $\varkappa_q$ that scales with $q$ in a power-law way
[see  Eq.~\eqref{kappa0}].
The components of the polarization operator scale then as follows \cite{lower-cr-D2, Doussal}:
 \be
\Pi_{\alpha\beta\gamma \theta}(\sigma=0)\propto
d_c \left( \frac{T}{q\varkappa_q}\right)^2 \propto \frac{1}{q^{2- 2\eta}},
\label{Pi-SCSA}
\ee
while the interaction constant scales in the universal region as
\be
Y_\mathbf q(\sigma=0)  = \frac{q^2\varkappa_q^2}{d_c T}  \propto q^{2- 2\eta}.
 \label{N-SCSA}
\ee

\begin{figure}[ht]
\centerline{\includegraphics[width=\linewidth]{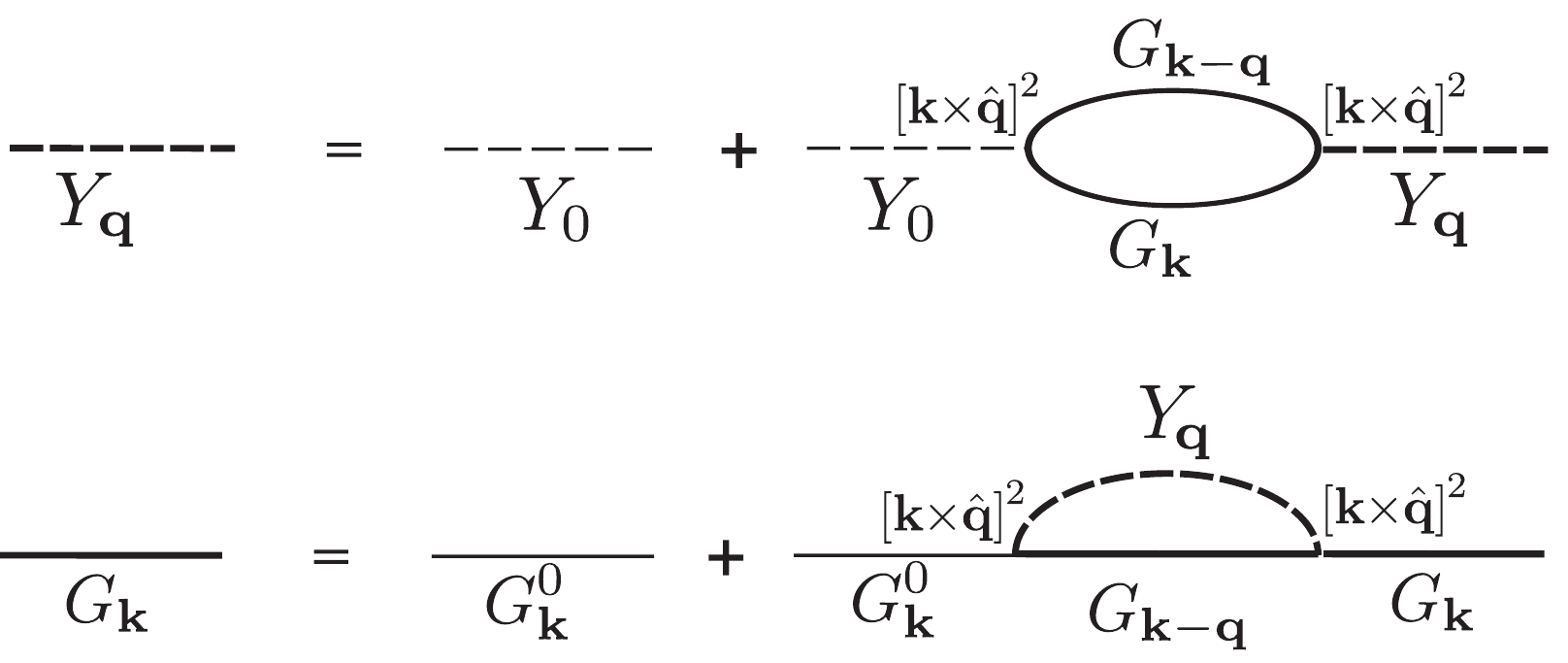}}
\caption{Self-consistent screening approximation: Graphical representation of the SCSA equations for the screened interaction $Y_\mathbf{q}$ and the propagator $G_\mathbf{k}$. The bare Green's function $G^0_\mathbf{k}$ is given by Eq. (\ref{harmonic}). Vertices are given by $[\mathbf k \times \mathbf {\hat q}]^2,$ where  $\hat q=\mathbf q/q$.}
\label{fig:SCSA}
\end{figure}

The critical scaling of the bending rigidity can be obtained within the SCSA scheme \cite{Doussal} by self-consistently solving the coupled equations for the self-energy and the RPA screened interaction with the simplest
polarization bubbles included, see Fig.~\ref{fig:SCSA}.
Within the SCSA, the critical exponent $\eta$ for a $2D$ membrane embedded into the space
of dimensionality $2+d_c$ is given by \cite{Doussal}
 \begin{eqnarray}
 \eta &=& \frac{4}{d_c+\sqrt{16-2d_c +d_c^2}} \nonumber \\[0.2cm]
 & \to &\left\{
 \begin{array}{cc}
\displaystyle \frac{2} {d_c}+\frac{1}{d_c^2},\qquad &\text{for} ~ d_c \to \infty, \\[0.4cm]
\displaystyle \! 1 -\frac{3 d_c}{16},\quad &\text{for} ~  d_c \to 0.
\end{array} \right.
 \label{eta-dc}
  \end{eqnarray}
Actually, only the asymptotics at  $d_c \to \infty $ is controlled by the small parameter
$1/d_c$. For an arbitrary $d_c$ (in particular, for the physical case $d_c=1$), there is no small parameter controlling the SCSA calculations.  Therefore, in the limit of large $d_c$, the  SCSA calculation yields correctly only the leading term $2/d_c$ in $\eta$, while an evaluation of higher-order corrections requires going beyond the SCSA.

The situation with the calculation of the PR is quite similar. In this case, in order to find the
absolute and differential PR, one should calculate functions $K_{\alpha}$ and $\Pi_{\alpha\beta}$ (which is found from a more general  polarization operator $\Pi_{\alpha\beta\gamma\theta}$), respectively.
There are two non-trivial problems  here: the inclusion of the external tension (in the non-linear regime) and an accurate account of finite-size effects (in the linear-response regime). As we are going to discuss below,
these complications do not allow one to find exact numerical values of $\nu$ and $\nu^{\rm diff}$, as this would require a resummation of all orders in $1/d_c$. Therefore, in what follows we will calculate the PR analytically only for $d_c\gg 1$ making use of the small parameter $\eta\ll 1$.

\begin{figure}[ht]
\centerline{\includegraphics[width=0.4\linewidth]{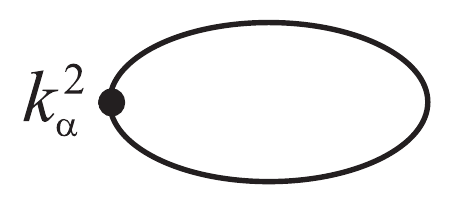}}
\caption{Diagram for function $K_{\alpha}$. The thick line denotes the propagator $G_\mathbf{k}$ at finite tension.
For calculating the terms in PR of zeroth and first orders in $1/d_c$, it is sufficient to use $G_\mathbf{k}$
defined in Fig.~\ref{fig:SCSA}.
}
\label{Fig-K}
\end{figure}

In order to calculate the terms in PR of zeroth and first orders in $1/d_c$ it is sufficient to evaluate the simplest diagram for $K_{\alpha}$ in Fig.~\ref{Fig-K} with the propagators $G_\mathbf{q}$ defined in Fig.~\ref{fig:SCSA}. This is similar to finding the leading-in-$1/d_c$ term in $\eta$ within the SCSA.
The corrections to the PR coming from diagrams that are not included in the set given by Figs.~\ref{fig:SCSA} and \ref{Fig-K} are quadratic-in-$1/d_c$. Importantly,
despite $\eta\ll 1$, the elastic coefficients are strongly renormalized at large $L$ and small $\sigma$. Thus, the evaluation of the PR at $d_c\gg 1$ does not assume the lowest-order perturbation theory
and requires a resummation of an infinite series of diagrams in terms of bare propagators (\ref{harmonic}).

Let us start with scaling estimates.
To this end, we take into account the tension by introducing the
term with the external tension into the denominator of Eq.~\eqref{G-SCSA}:
\begin{equation}
 G_{\mathbf q} =\frac{T}{\varkappa_q  q^4 + \sigma_x q_x^2 + \sigma_y q_y^2},
\end{equation}
which amounts to the replacement $\varkappa_0\to \varkappa_q$ in Eq.~\eqref{harmonic}.
This gives the following Green's functions
\be
G_{\mathbf q} =\frac{T}{\varkappa_q  q^4 + \sigma q_x^2} \quad\text{and} \quad
{\cal G}_{\mathbf q} =\frac{T}{\varkappa_q  q^4 + \sigma q^2},
\label{G-SCSA-1}
\ee
for the calculation of the absolute and differential PR, respectively.
These functions correctly describe the asymptotics of the true Green's functions in both limits
of large and small $q$. Specifically, they coincide with Eq.~\eqref{G-SCSA} for $q \gg \tilde q_\sigma$,
where \cite{my-hooke}
\be
\tilde q_\sigma  \simeq q_* \left (\frac{\sigma}{\sigma_*}\right)^{1/(2-\eta)}
\ee
is found from the condition $\varkappa_q q^2 \simeq \sigma $ and $\sigma_*$ is given by Eq.~\eqref{sigma-star}.
For $q \ll \tilde q_\sigma$, the tension terms in the denominators of the Green's functions dominate and one can neglect the  term $\varkappa_q q^4$. It is worth emphasizing  that the appearance of external tension $\sigma$ in the denominator of the Green's function of membrane with anharmonic coupling is a consequence of the corresponding Ward identity \cite{david2,lower-cr-D2,my-Ward}.
The approximation (\ref{G-SCSA-1}) corresponds to the neglect of the tension $\sigma$ in the self-energy of
the propagator $G_{\mathbf q}$. The status of this approximation will be discussed below.

The integrals entering $\delta K_\alpha$ and $\Pi_{\alpha \beta}$ are determined by $q\sim \tilde q_\sigma$.
Importantly, for $\sigma \ll \sigma_*$, the characteristic scale $ \tilde q_\sigma $ goes  beyond the inverse Ginzburg length $\tilde q_\sigma \ll q_*,$ so that  the membrane falls into the universal regime (see Fig.~\ref{Fig0}).
In particular, in the interval   $ \tilde q_\sigma \ll q\ll  q_*$, the components of the polarization operator
obey  the universal power-law scaling \eqref{Pi-SCSA} and saturate at  a value
$\sim (T/\tilde q_\sigma  \varkappa_{\tilde q_\sigma })^2$  for  $q \sim \tilde q_\sigma. $
Hence,
\be
\Pi_{\alpha \beta} \sim \left( \frac{T}{ q  \varkappa_{ q }} \right)^2_{q \simeq \tilde q_\sigma} \sim \frac{T}{\mu}\left ( \frac{\sigma_*}{\sigma} \right)^{1-\alpha},
\label{Pi-uni}
\ee
where $\alpha$ is given by Eq.~\eqref{alpha}.
Analogously,  estimating diagram shown in Fig.~\ref{Fig-K},  one finds
\be
\frac{\mu \delta K_\alpha}{\sigma} \sim  \left(\frac{\sigma_*}{\sigma}\right)^{1-\alpha}.
\label{K-uni}
\ee
As follows from Eqs.~\eqref{Pi-uni} and \eqref{K-uni},  for $\sigma \ll \sigma_*$  both absolute and differential PR are  fully determined by anomalous deformations and, therefore, universal.
They are given by  Eqs.~\eqref{nueff} and \eqref{nueff-abs}, respectively.
Hence,  the membrane exhibits universal elastic properties for  $\sigma \ll \sigma_*$.

There is also a lower bound on $\sigma$ for a membrane to show this universal PR. Indeed, we assumed above that the membrane has infinite size.   For a finite square-shaped membrane  with $   L_* \ll  L < \infty $,  one can neglect finite-size  effects provided that $\tilde q_\sigma \gg 1/L.$ The latter inequality  yields $\sigma \gg \sigma_L,$ where
\be
\sigma_L \sim \sigma_* \left (\frac{L_*}{L}\right)^{2-\eta} \ll \sigma_*.
 \label{sigma-L}
\ee
In the opposite limit,     $\sigma \ll \sigma_L,$   one can neglect tension terms in the denominator of  $G_{\mathbf  q}$ and    ${\cal G}_{\mathbf q}$ [see Eq.~\eqref{G-SCSA-1}] in the whole interval of $q \gtrsim 1/L.$  Then, the membrane shows linear response with respect to the external tension  and
\be
\nu=\nu^{\rm diff},\quad \text{for}\quad \sigma \ll \sigma_L.
\ee
A naive  approach to the analysis  of finite-size effects  is to  introduce the infrared   cut off $q \simeq 1/L $ into the  integrals determining $\delta K_\alpha$ and $\Pi_{\alpha\beta}.$
The PR in this regime still shows a certain universality, in the sense that it does not depend on microscopic details of the models. However, as we discuss in detail in Sec.~\ref{s5}, it is strongly sensitive to the BC which determine the system behavior at the scale $q\sim 1/L$. As a result, depending on BC,     the  PR at $\sigma=0$ can be either larger or smaller than its value in the universal regime,  $  \sigma_L<\sigma< \sigma_*$, see Fig.~\ref{Fig0}.

We return now to the regime of non-linear universality, $\sigma_L \ll \sigma \ll \sigma_*$, and emphasize the following important point.
Although the  approach described above captures correctly the scaling properties of the problem, it does not allow for the calculation of exact numerical values of  $\nu$ and $\nu^{\rm diff}$. Indeed, for calculation of the PR one needs to know the exact behavior of the Green's functions with the $\sigma$-dependent self-energy
$\Sigma(\mathbf{q};\sigma_x,\sigma_y)$,
\begin{eqnarray}
G_{\mathbf q}&=&\frac{T}{\varkappa_0q^4+\sigma_x q_x^2+\sigma_yq_y^2-\Sigma(\mathbf{q};\sigma_x,\sigma_y)}
\nonumber
\\
&\simeq&\frac{T}{\varkappa_qq^4+\sigma_x q_x^2+\sigma_yq_y^2-\delta\Sigma(\mathbf{q};\sigma_x,\sigma_y)},
\label{G-exact}
\end{eqnarray}
in the crossover region $q\simeq  \tilde q_\sigma$ [here  $\delta\Sigma(\mathbf{q};\sigma_x,\sigma_y)=\Sigma(\mathbf{q};\sigma_x,\sigma_y)-\Sigma(\mathbf{q};0,0)$ is the stress-induced correction to the self-energy].
In a generic situation (i.e., for $\eta \sim 1$), this behavior is complex and not known
even within the SCSA as defined in Fig. \ref{fig:SCSA}.
In particular, equations \eqref{G-SCSA-1}
for $G_{\mathbf q}$ and ${\cal G}_{\mathbf q}$ are only approximate (up to a ${\mathbf q}$-dependent factor of order unity) in the crossover region. The approximation becomes controllable only in the limit of $\eta \ll 1$. The point is that for small $\eta$ the bending rigidity grows very slowly and does not change essentially when $q$ varies  by a factor of the order of $2$.  Hence, in the leading order, one can use for calculations of PR the functions \eqref{G-SCSA-1} with the replacement of the scale-dependent $\varkappa_q $ with $\varkappa_{\tilde q_\sigma}$. However, already
linear-in-$\eta$ corrections to the PR are sensitive to the dependence of the self-energy on $\sigma$  and on the angle of $\mathbf q$. This dependence was neglected
within the approximation \eqref{G-SCSA-1} (which can be termed ``zero-$\sigma$ SCSA'') by setting $\delta\Sigma(\mathbf{q};\sigma_x,\sigma_y)=0$.  It is worth noticing  that, even at small $\eta$, the  bending rigidity at
$q \simeq  \tilde q_\sigma$  can be much larger than $\varkappa_0$ provided that the tension is sufficiently weak, $\ln (\sigma_*/\sigma) \gtrsim 1/\eta $.  The values of $\nu$ and $\nu^{\rm diff}$ in the small-$\eta$ (i.e., large-$d_c$) limit will be discussed in Sec.~\ref{small-eta}.

\subsection{Small $\eta$}
\label{small-eta}

For large dimensionality $d_c$ the calculation of the PR is controlled by a small parameter $\eta$. This allows us to develop a systematic expansion in $\eta$. We demonstrate this below by considering the differential and absolute PR.

\begin{figure}[ht!]
\centerline{\includegraphics[width=0.7\linewidth]{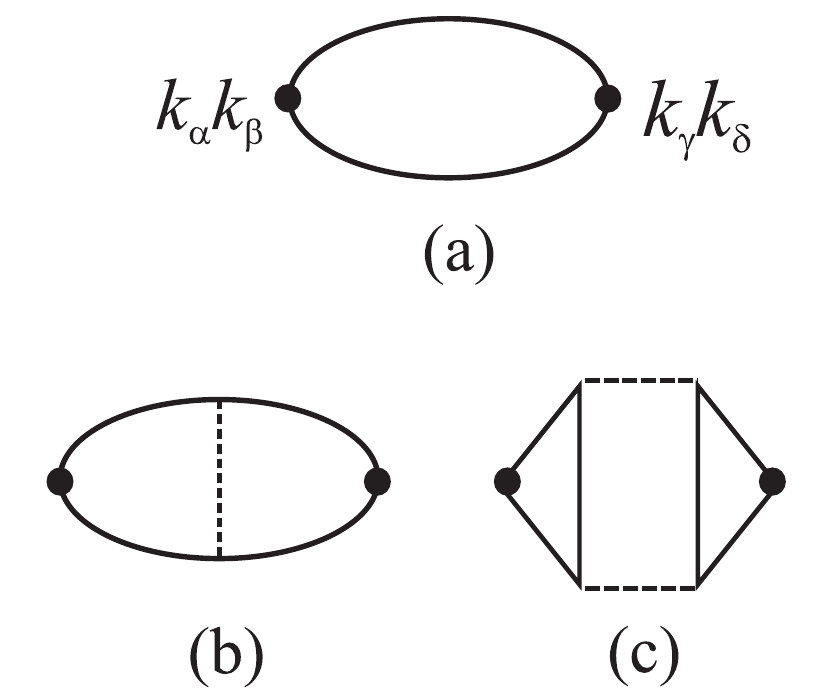}}
\caption{
Diagrams for polarization operator $\Pi_{\alpha\beta\gamma\theta}$
used for the calculation of the differential PR at $d_c\gg 1$.
Thick lines denote the renormalized propagator $G_\mathbf{k}$ and
dashed lines denote the renormalized interaction $Y_\mathbf{q}$ as defined in Fig.~\ref{fig:SCSA}.
Diagram (a) yields the leading term in the differential PR, while diagrams (b) and (c) provide the
$1/d_c$ correction. Diagram (c) has the same order as (b), since the smallness $1/d_c$ from the extra
dashed line is compensated by the factor $d_c$ coming from the extra loop. By virtue of Eq.~(\ref{pi alpha beta}),
the vertex-correction diagrams (b) and (c)
are obtained by differentiating the \textit{self-energy} of the propagator $G_\mathbf{q}$ in Fig. \ref{Fig-K}
with respect to $\sigma_\alpha$ and hence cannot be generated within the approximation (\ref{G-SCSA-1}).
}
\label{Fig-Pi}
\end{figure}

\subsubsection{Differential PR}

Let us consider diagrams in Fig.~\ref{Fig-Pi} for the polarization operator.
In the leading-in-$1/d_c$ approximation, the tensor  $\Pi^{\mathbf q}_ {\alpha\beta\gamma\delta}$ is
given by the diagram without vertex corrections shown in Fig.~\ref{Fig-Pi}(a),
where thick lines correspond to isotropic (in $q-$space)
functions  ${\cal G}_\mathbf q$ [see Eq.~\eqref{G-SCSA-1}].
Within this approximation, the polarization tensor is fully isotropic in the limit $q\to 0$:
  \be
   \Pi^{\mathbf{q} \to 0}_ {\alpha\beta\gamma\delta} =C(
   \delta_{ \alpha \beta}   \delta_{ \gamma \theta} +\delta_{ \alpha \gamma}\delta_{  \beta \theta} +\delta_{ \alpha \theta}\delta_{ \beta\gamma }),
  \ee
 where
 \be
 C=    \frac{d_c}{3}
 \lim_{q\to 0}
 \int \limits_{0}^{\infty}   k_\perp^4 {\cal G}_\mathbf k  {\cal G}_{\mathbf k-\mathbf q} \frac{d^2k}{(2\pi)^2}.
\label{C}
 \ee
Here ${\cal G}_\mathbf k$ is given by Eq.~\eqref{G-SCSA-1}  and $k_\perp^4 =|\mathbf k\times \mathbf q/q|^4$.
(This definition of the polarization operator  differs   by a  numerical coefficient from the one  used in Ref.~\cite{my-crump}.)
Hence, within this $d_c\to\infty$ approximation
\begin{align}
 &\Pi_{xx}=3\Pi_{xy}=  3C,
\\
& \Pi_+=\Pi_-=2C
\label{Pi-one-loop}
\end{align}
Substituting this in  Eq.~\eqref{nueff}, we obtain $\nu^{\rm diff}=-1/3$, which is exactly the result of
Ref.~\cite{Doussal}. It is worth noting that this number
can be obtained in a straightforward way  without actual calculation of the integral over $k$ in Eq.~\eqref{C}.
Indeed, in the absence of vertex corrections, $\Pi_{xx} \propto \langle n_x^4 \rangle $ and
$\Pi_{xy} \propto \langle n_x^2n_y^2 \rangle$, with the same coefficient.
Here $\mathbf n=\mathbf k/k$ and $\langle \cdots \rangle$ stands for the angle averaging.
Then, we get  $\langle n_x^4 \rangle=3/8,$ and   $\langle n_x^2 n_y^2 \rangle=1/8,$ and finally obtain
 \be
 \nu^{\rm diff}=-\frac{\langle n_x^2 n_y^2 \rangle}{\langle n_x^4 \rangle}=-\frac{1}{3}.
 \label{1/3}
 \ee

Importantly, this result is not valid when vertex corrections are included in the polarization bubble,
such as in Figs.~\ref{Fig-Pi}(b) and (c).
A general structure of the polarization operator is \cite{lower-cr-D2}
\be
\Pi_{\alpha \beta, \gamma \theta}^{\mathbf{q}\to 0}=C_1 \delta_{ \alpha \beta}   \delta_{ \gamma \theta} +C_2 (\delta_{ \alpha \gamma}\delta_{  \beta \theta} +\delta_{ \alpha \theta}\delta_{ \beta\gamma }),
\ee
with $C_1\neq C_2$.  Hence,  in the general case,
\be
\Pi_{xx}=C_1+2 C_2, \quad \quad \Pi_{xy}=C_1.
\label{pixx-pixy-C-D}
\ee
A direct analysis of diagrams Fig.~\ref{Fig-Pi}(b) and (c) shows that the condition $C_1=C_2$ indeed fails  when the vertex corrections are included. Consequently, conditions   $\Pi_{xx}=3\Pi_{xy}$ and $\Pi_+=\Pi_-$,  also fail.
In a general case,  $\Pi_{xx}-3\Pi_{xy}=2(C_2-C_1).$
Hence, in a generic membrane with $\eta \sim 1$ the differential PR is not  equal to $-1/3$.  This applies, in particular, to physical membranes in a three-dimensional space, in which case $d_c=1$ and $\eta \approx 0.7 \div 0.8$.

Let us discuss this point in more detail. A scaling analysis of the diagrams  in Fig.~\ref{Fig-Pi} shows that for
 $ q_\sigma \ll q \ll q_*$ the  components of the  polarization operator increase with decreasing $q$   as
 $1/q^{2-2\eta}$:
  \be
 \Pi_{xxxx} =C_{xx} \frac{T^2} { (q \varkappa_q) ^{2}},\quad \Pi_{xxyy} =C_{xy} \frac{T^2} { (q \varkappa_q) ^{2}},
 \label{eq:67}
  \ee
   where   $C_{xx}$ and  $C_{xy}$ are numerical coefficients. The PR is given in terms of these coefficients as
  \be
  \nu^{\rm diff }=-\frac {C_{xy}}{C_{xx}}.
  \label{eq:67a}
  \ee
To the leading order (zeroth order in $\eta$), the PR is determined by the diagram of Fig.~\ref{Fig-Pi}(a), yielding
$\nu^{\rm diff} = -1/3$, as discussed above.
  Corrections of the first order in  $\eta$ come from the diagrams  \ref{Fig-Pi}(b) and   \ref{Fig-Pi}(c).
Indeed, each interaction line gives a factor $1/d_c \sim \eta$,  see Eq.~(\ref{N-SCSA}), so that the diagram \ref{Fig-Pi}(b) yields a correction of the order $\eta$. The diagram \ref{Fig-Pi}(c), although of second order in interaction, contains an additional polarization loop that gives a factor $d_c \sim 1/\eta$, and thus contributes to the order $\eta$ along with the diagram \ref{Fig-Pi}(b).

The calculation of the numerical coefficient resulting from the  diagrams \ref{Fig-Pi}(b) and (c) is lengthy and will be presented elsewhere \cite{my-Ward}. The result reads
\be
\nu^{\rm diff} =  -\frac{1}{3} + 0.008 ~\eta + O(\eta^2).
\label{nu-diff-uni}
\ee
The correction proportional to $\eta$ is due to a difference between the polarization operators $\Pi_+$ and $\Pi_-,$
which emerges in the order $\eta$:
\be
\frac{\Pi_+ - \Pi_- }{ \Pi_+ } \simeq 0.018 \eta.
\ee

Thus, the  exact value  of $\nu^{\rm diff}$ is a universal function of  $\eta$ which can be obtained in a controllable way by an expansion over powers of $\eta$.
The value $-1/3$ is obtained only in the limit $\eta \to 0$, i.e., at $d_c \to \infty$.   For a finite $d_c$, vertex corrections lead to appearance of a non-zero value of $(\Pi_+-\Pi_-)/\Pi_+$ which is absent in the lowest order and yields corrections to $\nu^{\rm diff}$.

\subsubsection{Absolute PR}

Let us now consider the absolute PR. Within the approximation (\ref{G-SCSA-1}), the anomalous deformation reads
\BEA
&&  K_\beta(\sigma,0)=d_c T \int \frac{d^2 q}{ (2\pi)^2}\frac{q_\beta^2}{\varkappa_q q^4+ \sigma_{\varphi} q^2 }
\nonumber
\\
&& =K(0)- \delta K_\beta(\sigma,0).
\EEA
All stress dependence is encoded in the function
\be
\delta K_\beta(\sigma,0) =
\frac{d_c T}{2\pi}\left\langle \int\limits_{0}^\infty  \frac{d q}{  q}\frac{n_\beta^2 \sigma_{\varphi} }{\varkappa_q  \left(\varkappa_q q^2+ \sigma_{\varphi} \right) } \right\rangle_\varphi.
\label{deltaq}
\ee
Here
\be
\sigma_\varphi= \sigma \cos^2\varphi,
\label{sigma-fi}
\ee
 and $\langle \cdots \rangle_\varphi$ stands for averaging over $\varphi$.
   It is worth emphasizing that in contrast to $\nu^{\rm diff},$ which represents a linear response to a small anisotropic stress on top of a large isotropic one, the absolute PR describes an essentially non-linear response. This is clearly reflected in Eq.~\eqref{deltaq}, where the applied stress $\sigma_\varphi$ enters  both the numerator and the denominator.
 The integral \eqref{deltaq} is determined by momenta   $q \sim \tilde q_\sigma .$  For $\sigma  \ll \sigma_*$
   such $q$ are located in the region of the anomalous elasticity $q \ll q_*$, where the renormalized bending rigidity $\varkappa_q$ scales with $q$ in a power-law way. As a result, $\delta K_\beta$  turn out  to be power-law functions of $\sigma$.  For   $\sigma \gg \sigma_*$,   the bending rigidity $\varkappa_q$ is approximately given by its bare value $\varkappa$, so  that $\delta K_\beta$  grows as $\ln \sigma$.  Thus, we get from Eq.~\eqref{deltaq}
   \be
   \frac{Y_0 \delta K_\beta }{2\sigma}=  \left\{ \begin{array}{lcl}
                                         A_0\left(\sigma_*/\sigma\right)^{1-\alpha} \left \langle  n_x^{2\alpha} n_\beta^2 \right \rangle_\varphi,\quad\text{for}~\sigma\ll\sigma_*,
                                          \\
                                          A_1\left(\sigma_*/\sigma\right) \ln\left(\sigma_*/\sigma\right),\qquad  \hspace{0.45cm} \text{for}~\sigma\gg\sigma_*,
                                       \end{array}\right.
   \label{dK}
   \ee
   where $A_0$ and $A_1$ are numerical coefficients of order unity and $\alpha$ is given by Eq.~\eqref{alpha}.
   (We  remind the reader that in  all estimates we   assume that all bare elastic constants are of the same order  and absorb the corresponding dimensionless ratios in prefactors of order unity.)

 It is seen from Eqs.~\eqref{nu-absolute} and \eqref{dK} that for $\sigma \gg \sigma_*$ the Young modulus and the PR are given, to the leading order, by their bare values. On the other hand, for $\sigma \ll \sigma_*$ anomalous terms dominate and we find
   \BEA
   \label{Y}
   Y &\sim&  Y_0(\sigma/\sigma_*)^{1-\alpha}
   \EEA
   and the following result for the PR,
   \BEA
   \nu &=&\nu_{\rm min} = - \frac{\left \langle \cos^{2\alpha}\varphi ~ \sin^2 \varphi \right \rangle_\varphi}{\left \langle \cos^{2\alpha+2}\varphi \right \rangle_\varphi}=-\frac{1}{1+2\alpha} \nonumber
   \\
   &=&  -\frac{2-\eta}{2+\eta}.
      \label{nu}
   \EEA
Hence,  the Young modulus $Y$ is suppressed due to the softening of membrane by thermal fluctuations, while $\nu$ equals to a certain universal value,
    $\nu_{\rm min }$,
which is determined solely by the critical index $\eta$ (i.e., by the dimensionality $d_c$).   In full analogy with the differential PR, the result \eqref{nu} for the absolute PR is strictly valid  only at $\eta=0$, yielding in this limit
$\nu_{\rm min}=-1$. For a generic $\eta \sim 1$ (and, in particular, for the physically relevant case $d_c=1$) it constitutes an uncontrollable approximation.  In order to find a correction of the first order in $\eta$, one should take into account, in  addition to the correction $\sim\eta$ entering Eq.~\eqref{nu}, the deviation of the propagator
$G_\mathbf q$ from its approximate form (\ref{G-SCSA-1}). The difference stems from the dependence of the self-energy on
$\sigma$.
The resulting expansion of  $\nu$   up to the first-order term in $\eta$ reads
\be
\nu_{\rm min}=-1+ (1 + C_{\Sigma})\eta +O(\eta^2) \,,
\label{nu-m}
\ee
where $C_{\Sigma}$ is the contribution to the coefficient of the $\eta$ correction resulting from the stress dependence of self-energy. The evaluation of the numerical coefficient $C_{\Sigma}$ is very tedious and is
postponed to a forthcoming publication. Importantly, for an anisotropic tension, the exact self-energy in Eq. \eqref{G-exact} depends on the angle of $\mathbf q$, so that the effect of anharmonic interaction at $q \simeq  \tilde q_\sigma$ is not fully captured by the $|q|$-dependent function $\varkappa_q$.

Substituting Eq.~\eqref{dK} into Eq.~\eqref{nu-absolute}, one can find subleading correction to Eq.~\eqref{nu}:
   \be
   \nu_{\rm min} (\sigma) \approx  \nu_{\rm min }+ A_2 (\sigma/\sigma_*)^{1-\alpha} ,\quad{\rm for}\quad \sigma_{L} \ll \sigma \ll \sigma_*,
   \label{nu-with-corr}
   \ee
where $A_2 \sim 1$ is a positive numerical coefficient.
The lower boundary $\sigma_L$ of the region of validity of Eq.~\eqref{nu-with-corr} is determined by the system size $L$ [see Eq.~\eqref{sigma-L}]
providing the infrared cutoff to Eq.~\eqref{deltaq}.  In the limit of an infinite system, $L = \infty$,
we have $\sigma_L = 0$, so that Eq.~\eqref{nu-with-corr} is applicable down to arbitrarily weak stresses.

Two comments are in order  before closing this subsection. First, comparing Eq.~\eqref{nu-m} with Eq.~\eqref{nueff-abs}, we see that  $\delta K_+ \neq \delta K_- $ even in the limit $\eta \to 0.$    As seen from Eq.~\eqref{dK}, this happens because in this limit  (which implies  also $\alpha \to 0$),
$\delta K_\beta  \propto \langle n_\beta^2\rangle.$  Hence, anomalous deformations  in $x$ and $y$ direction coincide:
$\delta K_x=\delta K_y$ and, consequently,  $\delta K_-\equiv 0$  [see Eq.~\eqref{dKpm}]  for  $\eta=0.$ Then we find from Eq.~\eqref{nueff-abs}  $\nu \to -1$, in agreement with Eq.~\eqref{nu-m}.

The second comment concerns the  generalization of the  above results   obtained under condition of   the  uniaxial stress  ($\sigma_x=\sigma,~\sigma_y=0$)  to the case of more general deformations. This should be done with caution. In particular,
the uniaxial modulus $C_{11}$---which is one of most conventional characteristics of an elastic media---is ill-defined for a membrane
with negative PR. Indeed,  by definition, $C_{11}$  corresponds to
 $ \varepsilon_y= 0.$ The latter condition prevents expansion in
$y-$direction and therefore should lead to a transverse wrinkling
instability. A detailed study of this instability is
an interesting prospect for future work.

\subsection{$\eta \to 1$} \label{eta-1}

As we have shown above, the large-$d_c$ limit (which corresponds to small $\eta$) allows one to get a controllable approximation (and, in principle, also an expansion in powers of $\eta \sim 1/d_c$) for the PR.  It is natural to ask whether the opposite  limit of small $d_c$ can be used to develop a complementary approximation. The limiting value $d_c=0$ corresponds to a 2D membrane embedded into 2D space. Evidently, because of absence of out-of-plane modes, the   anomalous deformations, which are proportional to number of transverse modes,   are  exactly equal to zero in this limit,   $ K_\beta \propto d_c  = 0$ for $d_c = 0$. Hence, in-plane moduli are  not renormalized by  anomalous elasticity in this limit, so that the membrane should  obey conventional Hooke's law [Eqs.~\eqref{eps-x} and \eqref{eps-y} with $\delta K_\beta=0$]. This corresponds to the index $\alpha=1$, and thus, according to Eq.~(\ref{alpha}), $\eta = 1$.    In this situation, both the absolute and  differential  PR are equal to the non-universal material value $\nu_0$:
\be
\nu=\nu_{\rm dis}=\nu_0\quad \text{for} \quad\eta=1.
\ee

Let us now turn to the range of small but non-zero dimensionality $d_c$.  In this case,  $\eta$ is not exactly equal to unity  but is close to it, $1-\eta \sim d_c\ll 1$.  The polarization operator then behaves logarithmically,
\be
\Pi_{\alpha\beta\gamma\theta} \propto   d_c  \frac{T}{\mu}   \ln\left (\frac{q_*}{q}\right),
\ee
in an exponentially   wide interval of $q$:
  \be
  \tilde q_* < q < q_* \,,
  \label{interval}
  \ee
 where
 \be
 \tilde q_* \sim q_* \exp (-1/d_c),
 \ee
 is inversely proportional to  the spatial scale $\tilde L_*$ at which the anomalous elasticity becomes fully developed:
 $\tilde L_*\sim 1/\tilde q_*$.  Within the interval    \eqref{interval}, the anomalous deformations can be treated perturbatively, so that PR (both absolute and differential) remains close to $\nu_0$. Only for exponentially small wave vectors,  $q \ll \tilde q_*$ (or, equivalently, for exponentially large size, $L \gg  \tilde L_*$),  the membrane falls into the universal regime, so that   anomalous deformations  scale in a  power-law  way with $q$.  In particular, the polarization operator then scales in
 accordance with Eq. \eqref{eq:67}.
The differential  PR takes a universal value
which can be determined from Eq. \eqref{eq:67a}.

Unfortunately, at this stage, we are not able to present a controllable scheme for the evaluation of this universal value of $\nu^{\rm diff}$ for small $d_c$.
The result   $\nu^{\rm diff}=-1/3$ given by the diagram of Fig.~\ref{Fig-Pi}(a) is strongly modified by higher-order corrections. The only essential simplification is the smallness of diagrams containing more then one full-line loop, such as the diagram shown in Fig.~\ref{Fig-Pi}(c). Such diagrams can be neglected, since each loop gives an additional small factor $d_c.$  On the other hand, one can check that  all high-order diagrams containing a single full-line loop do not have any additional small factors $\sim d_c$ and, therefore, should be taken into account along with diagram in
Fig.~\ref{Fig-Pi}(a). The evaluation of the numerical value of $\nu^{\rm diff}$  in this regime thus requires a summation of an infinite set of diagrams, which remains a challenging problem for future research.  A similar conclusion holds for the absolute PR for $q\ll \tilde q_*$.  Via the same token, one can check that  the value of  the coefficient in the first order of the expansion of the exponent $\eta$  over $d_c$ is  modified  in an uncontrollable way by higher-order terms  as compared to  the value $-3/16$ in the lower line of Eq.~\eqref{eta-dc} (cf. a discussion in Ref.~\cite{Gazit1}).

To summarize, we find that for small $d_c$ (i.e., $\eta$  close to unity) both the absolute and differential PR remain close to the non-universal material parameter $\nu_0$  within the broad interval of system sizes  determined by Eq.~ \eqref{interval} but eventually flow to still unknown values  for the  exponentially large systems,
$L \gg  1/\tilde q_*$.

\section{Finite-size effects}
\label{s5}

In this Section, we analyze the effect of a finite size of a membrane on the PR. Consider, for example, the absolute PR. Within the apprximation (\ref{G-SCSA-1}), the finite size $L$  of the system determines the infrared cutoff in the integral in Eqs.~\eqref{deltaq}.   As a result, Eqs.~\eqref{Y} and \eqref{nu} become invalid at low  stress $\sigma < \sigma_L$, where $\sigma_L$ is given by Eq.~\eqref{sigma-L}.
Let us consider a square-shaped  sample with the  size $L \gg L_*$  and  assume that the membrane is stretched  by  small  tensions:
  \be
   \sigma_x \ll  \sigma_L,  \quad \sigma_y  \ll  \sigma_L .
   \ee
We will demonstrate that in this case  the value of  the PR
strongly depends on the boundary conditions.  This should be taken into account when one compares results of analytical, numerical, and experimental evaluation of PR. The importance of the BC for numerical simulations of the PR was recently pointed out in Ref.~\cite{Ulissi2016}.

\subsection{Poisson ratio}

In the regime that we are considering, thermodynamic fluctuations of the strain are relatively strong, as will be discussed below. The PR is defined as
\be
\nu =- \frac{\langle \varepsilon_y \rangle  }{  \langle \varepsilon_x \rangle },\quad \nu^{\rm diff } =- \frac{\langle \delta \varepsilon_y \rangle  }{ \langle \delta \varepsilon_x \rangle },
\ee
where $\langle \ldots \rangle$ denotes the thermodynamic averaging. This definition corresponds to  the diagrammatic approach described in the previous sections.

For a finite system, the integration in the   equation  \eqref{deltaq} for the anomalous deformations should be replaced with the summation.  Minimal $q$  entering the sum     is limited by $\sim 1/L.$  Then,  one can neglect $\sigma _\varphi=\sigma_x\cos^2 \varphi+\sigma_y\sin^2\varphi$ in the denominator of the integrand. As a consequence, $\delta K_\beta$ becomes a  linear function of  $\sigma_x$  and $\sigma_y$ and the absolute and differential responses coincide:
 \be
 \nu=\nu^{\rm diff} \quad \text{ for} \quad \sigma \ll \sigma_L,
 \ee
which is a manifestation of the fact that we are in the linear-response regime with respect to the external stress.
For a not too small system, $L\gg L_*$,  anomalous deformations dominate over the conventional ones, and $\varepsilon_\beta \simeq \delta K_\beta/2.$
The scaling  of $\delta K_\beta$  can be understood within the approximation (\ref{G-SCSA-1}) (one can check  that high-order corrections do not change this scaling).

Assuming a symmetry between the $x$ and $y$ axes (which implies a square shape of the sample), one can write the  balance equations, which have the form analogous to  Eqs.~\eqref{dif-eps-x} and \eqref{dif-eps-y},
\BEA
 \varepsilon_x&\simeq&   \frac{1}{Y_{L}} (\sigma_x -\nu \sigma_y),
 \label{1anom}
 \\
 \label{2anom}
 \varepsilon_y&\simeq &\frac{1}{Y_{L}} (-\nu \sigma_x+\sigma_y),
 \EEA
with the  size-dependent Young modulus
 \be
 Y_{L}\simeq  Y_0 \left (\frac{L_*}{L}\right)^{2-2\eta}.
 \label{Yeff}
 \ee
  In Eq.~(\ref{Yeff}) we have used the fact that in the considered regime the system size $L$ provides the infrared cutoff for the anomalous scaling of the elastic modulus.   Equations ~\eqref{1anom} and \eqref{2anom} represent the general form of  the balance  equations in the linear regime ($\sigma \ll \sigma_L$) of the finite membrane with  large size ($L \gg  L_*$).     The numerical coefficient in Eq.~\eqref{Yeff}  depends on microscopic details of the system at the ultraviolet scale and on the precise definition of $L_*$.    On the other hand, the PR $\nu$ is not sensitive  to the material-dependent (ultraviolet) physics and is universal in this sense. However, $\nu$ does depend on BC, as we discuss in detail below. Furthermore, if one considers a sample of an arbitrary aspect ratio, this will also influence $\nu$. For definiteness, we focus on a square geometry of the sample below.

 Before turning to the analysis of the PR for various BC, we notice that other  elastic coefficients  are also size-dependent  and are  related to $Y_L$ and $\nu$ by conventional equations of the elasticity theory. In particular, the bulk and uniaxial moduli are proportional to $Y_L$:
 \be
 B=\frac{Y_L}{2(1-\nu)},\quad C_{11}=\frac{Y_L}{(1-\nu^2)}.
 \label{B,C11}
 \ee
We emphasize  again that Eqs.~\eqref{1anom},\eqref{2anom},\eqref{Yeff}, and \eqref{B,C11} are valid for arbitrary BC independently from the microscopic model on the ultraviolet scale.   All the information on the non-universal (material-dependent) ultraviolet physics is contained in the parameters $Y_0$ and $L_*$  in Eq.~\eqref{Yeff}.

To calculate $\nu$ in a controllable way, we consider the limit of $d_c \to \infty$, i.e., $\eta \to 0$.
Naively, one could attempt to get the result by introducing an infrared cut-off  $q_{\rm in} \sim 1/L $ in the
integral in Eq.~\eqref{deltaq}.  Neglecting  $\sigma _\varphi$ in the denominator of the integrand in Eqs.~\eqref{deltaq}, we would then find that the main contribution to the integrals comes then from the lower limit $q_{\rm in}$, yielding $\nu=-1/3$ independently of the exact value   $q_{\rm in}$. Analyzing this naive calculation,
we observe that the value 1/3 can be traced back to the ratio of two angular averages, $- \nu = \left \langle  n_x^2n_y^2 \right\rangle_\varphi / \left \langle  n_x^4 \right\rangle_\varphi = 1/3$. The origin of this value is exactly
the same as that  of the value $\nu^{\rm diff} = -1/3$ in the regime of non-linear universal elasticity $\sigma_L \ll \sigma \ll \sigma_*$, see Eq.~\eqref{1/3}.  One might thus come to a conclusion that the result $-1/3$ remains valid for the differential PR of the $\eta\to 0$ problem also for $\sigma \ll \sigma_L$. However, this conclusion is incorrect. Contrary to the regime $\sigma \gg \sigma_L$ [in which Eq.~\eqref{1/3} holds], the replacement of  summation by  integration in the regime $\sigma \ll \sigma_L$ is not justified.

We use  Eq.~\eqref{nueff}, where  $\Pi_{\alpha \beta}$ are given by the diagram in Fig.~\ref{Fig-Pi}(a) (which is the dominant contribution in the limit $\eta \to 0$). Taking into account the discreteness of momenta in course of evaluation of  $\Pi_{\alpha \beta}$, we find
 \be
 \nu=\nu^ {\rm diff} \simeq -\frac{\Pi_{xy} }{ \Pi_{xx}}=
-
 \frac{\displaystyle
 \sum \limits_{\mathbf q} \frac{q_x^2 q_y^2}{ q^{8}}
  }
  { \displaystyle
  \sum \limits_{\mathbf q} \frac{q_x^4}{ q^{8}}
  }.
  \label{double-sum}
 \ee
 To define unambiguously the sums in the numerator and denominator of Eq.~\eqref{double-sum}, one has to specify the BC.
  For the simplest case of  periodic BC, $h(x+L,y)=h(x,y+L)=h(x,y),$
  quantized wave vectors are given by   $q_x=2\pi n/L, q_y=2\pi m/L. $ The point $n=m=0$ should be excluded from summation both in the numerator and denominator of Eq.~\eqref{double-sum}. We obtain
  \be
  \nu_{\rm per}=-\frac{\displaystyle 4 \sum \limits_{n=1}^\infty \sum \limits_{m=1}^\infty \frac{n^2m^2}{(n^2+m^2)^{4}}}
 { \displaystyle
 4 \sum \limits_{n=1}^\infty \sum \limits_{m=1}^\infty \frac{n^4}{(n^2+m^2)^{4}} +2 \sum \limits_{n=1}^\infty \frac{n^4}{(n^2)^{4}}
 }=-0.135.
 \label{per}
 \ee
 This result  can be straightforwardly generalized to the cases of  free  and zero BC, $\p_x h=\p_yh=0,$ and $h=0,$ respectively:
  \be
  \nu_{\rm free}=-\frac{\displaystyle  \sum \limits_{n=1}^\infty \sum \limits_{m=1}^\infty \frac{n^2m^2}{(n^2+m^2)^{4}}}
 { \displaystyle
  \sum \limits_{n=1}^\infty \sum \limits_{m=1}^\infty \frac{n^4}{(n^2+m^2)^{4}} + \sum \limits_{n=1}^\infty \frac{n^4}{(n^2)^{4}}
 }=-0.075
 \label{free}
 \ee
and
 \be
 \nu_{\rm zero}=-\frac{\displaystyle  \sum \limits_{n=1}^\infty \sum \limits_{m=1}^\infty \frac{n^2m^2}{(n^2+m^2)^{4}}}
 { \displaystyle
  \sum \limits_{n=1}^\infty \sum \limits_{m=1}^\infty \frac{n^4}{(n^2+m^2)^{4}}.
 }=-0.735
 \label{zero}
 \ee

Inspecting Eqs.~\eqref{per}, \eqref{free}, and \eqref{zero}, we see that $\nu$ in the linear-response regime $\sigma \ll \sigma_L$ can change dramatically (by an order of magnitude) depending on BC. Another interesting observation is that, for some types of BC, the differential PR is a non-monotonous function of the stress. Indeed, comparing Eqs.~\eqref{per}, \eqref{free}, and \eqref{zero} with Eq.~\eqref{nu-diff-uni}, we see that this is the case for the  periodic and free BC, because $|\nu_{\rm per}| <1/3$ and $|\nu_{\rm free}| <1/3$. Such a situation is shown as BC 1 in Fig.~\ref{Fig0}. On the other hand, for zero BC,  $|\nu_{\rm zero}| >1/3$ and $\nu^{\rm diff}$ monotonously grows with increasing $\sigma$, as shown by the  BC 2 curve  in Fig.~\ref{Fig0}. We remind the reader that the values \eqref{per}, \eqref{free}, and \eqref{zero} are valid in the $\eta \to 0$ limit. In the case of a finite $\eta$ (e.g., for a physical situation of $d_c=1$ with $\eta \simeq 0.7$), the numerical values will be different. However, the strong variation of $\nu$ with BC will definitely  persist. Furthermore, it is highly plausible that the dependence of $\nu^{\rm diff}$ on $\sigma$ will remain non-monotonous for some of BC.

\subsection{Fluctuations of strain and stress}

Here we estimate fluctuations of the strain and stress of a finite-size system. First, we notice that for fixed  configuration  of out-of-plane deformation field, $\mathbf h=\mathbf h(\mathbf r),$ the  stresses $\sigma_x$ and $\sigma_y$ are given by Eq.~\eqref{tensor-hooke} with only spatial averaging but without Gibbs averaging:  $K_\alpha \to K_\alpha^0$, cf. Eqs.~\eqref{K-ave} and \eqref{K-L}.  Then, we express deformations,  $\varepsilon_\alpha=  (\xi_\alpha^2 -\xi_0^2)/2\approx \xi_\alpha -1$   (neglecting the  difference between $\xi_0$ and unity) as
\be
\varepsilon_\alpha =- \frac{K_\alpha^0}{2} +\hat M^{-1}_{\alpha\beta} ~\sigma_\beta.
\label{eps-sig}
\ee

In the universal region,  $L\gg L_*,$ one can neglect  the second term in Eq.~\eqref{eps-sig}, which yields
for the distribution function  of strain
\be
f_{\alpha} (\varepsilon_\alpha) =\left \langle \delta\left (\varepsilon_\alpha   +{ \overline{(\p_\alpha \mathbf h)^2}}/{2} \right ) \right \rangle.
\label{dist}
\ee
Here $\langle \dots \rangle$ is the Gibbs averaging for fixed $
\sigma_x $ and $\sigma_y$, which is taken with the functional
\be
E= \frac{1}{2} \sum\limits_\mathbf q  (\sigma_\alpha  q_\alpha^2    +\varkappa_q q^4 )  |\mathbf h_\mathbf q|^2.
\label{Esigma}
\ee
which corresponds to the approximation (\ref{G-SCSA-1}) for the Green's functions.
 Performing this averaging,
   we find
 \begin{eqnarray}
f_{\alpha} (\varepsilon_\alpha) &=&
\int\frac{dz}{2\pi} e^{iz\varepsilon_\alpha } \label{dist1}\\
&\times&\!\prod \limits_\mathbf q\!
\left(\!\frac{\varkappa_ q q^4+\sum \limits_\beta \sigma_\beta  q_\beta^2 }
{-i z T q_\alpha^2/L^2 +\varkappa_ q q^4 +\sum \limits_\beta \sigma_\beta  q_\beta^2   }\!\!\right)^{\!d_c/2}.
\nonumber
\end{eqnarray}
 Here we take into account  only fluctuations of deformations caused by out-of-plane modes. One can check that the effect of in-plane modes is  parametrically smaller provided that $L$  is much larger than $\sqrt{\varkappa/\mu}. $  The latter scale is of the order of the lattice constant and can be considered as an ultraviolet cut-off of the theory \cite{my-quantum}. Hence, contribution of in-plane modes can be safely neglected.

The   integrand in Eq.~\eqref{dist1}  shows  a simple  pole  structure as  a function of $z$  thus allowing for simple analytical calculations of the  moments of the deformation distribution.  Evidently, $\int f_\alpha (\varepsilon_\alpha)  \  d \varepsilon_\alpha=1.$   Calculating next $\langle \varepsilon_\alpha \rangle  =\int f_\alpha (\varepsilon_\alpha)  \varepsilon_\alpha  d \varepsilon_\alpha$, we recover Eq.~\eqref{Kk} with the Green function
$G_\mathbf q =T/(\varkappa_q q^4+ \sum \limits_\beta \sigma_\beta  q_\beta^2 ).$
  A  direct calculation of  the fluctuation amplitude by using Eq.~\eqref{dist1} yields
        \be
  \Delta \varepsilon =  \sqrt {\langle  (\varepsilon_\alpha -  \langle  \varepsilon_\alpha     \rangle)^2   \rangle}
   =\sqrt{\frac{d_c T^2}{2 L^4}\sum \limits_\mathbf q  \frac{q_\alpha^4}{(\varkappa_q q^4+ \sum \limits_\beta \sigma_\beta  q_\beta^2)^2 }},
     \label{fluct-eps0}
   \ee
    thus leading to the
    the following result
   \be
  \Delta \varepsilon^2
   \sim d_c~ \left \{ \begin{array}{c}
({ T}/{\varkappa_{1/L}})^2,  \quad \text{for}\;\; \sigma \ll \sigma_L, \\[0.3cm]
  (T^2/\varkappa_{\tilde q_\sigma} \varkappa_{1/L})          (\sigma_L/\sigma),   \quad \text{for}\;\; \sigma \gg \sigma_L.
        \end{array}
        \right.
  \label{fluct-eps}
   \ee
      For simplicity,  in Eq.~\eqref{fluct-eps} we considered isotropic case  $\sigma_x =\sigma_y = \sigma.$
    Hence, fluctuation are suppressed for $\sigma \gg \sigma_L$ and become independent on $\sigma$ for $\sigma \ll \sigma_L$.

In the limit $d_c \gg 1$, the distribution of strain $f_\alpha$  represents a Gaussian peak  centered at $\langle \varepsilon_\alpha \rangle$ with a width  given by Eq.~\eqref{fluct-eps0}. On the other hand, for $d_c \sim 1$,  the distribution function $f_\alpha$ becomes essentially asymmetric at $\sigma \ll \sigma_L$.

Equation \eqref{fluct-eps} yields fluctuation of strain $\varepsilon$ at fixed stress $\sigma$.  It can be used to determine fluctuations of $\sigma$ at fixed strain. The result reads
 \BEA
   && \Delta \sigma \sim \frac{\Delta \varepsilon}{ |\p  \langle (\nabla \mathbf h )^2 \rangle /\p \sigma|  }
  \sim \frac{\Delta \varepsilon}{ |\p  \langle \varepsilon  \rangle /\p \sigma|  }
  \nonumber
  \\
   && \sim \frac{1}{\sqrt{d_c}}\left \{ \begin{array}{c}
          \sigma_L,  \quad \text{for}\;\; \sigma \ll \sigma_L, \\ \\
          \sqrt{(\varkappa_{\tilde q_{\sigma}}/\varkappa_{1/L})}~ \sqrt{\sigma_L \sigma },   \quad \text{for}\;\; \sigma \gg \sigma_L.
        \end{array}
        \right.
        \label{DeltaSigma}
   \EEA
The factor $1/\sqrt{d_c}$ suppressing the fluctuations in the limit $d_c \to \infty$ originates from the self-averaging
of the fluctuations for a large number of out-of-plane modes.
As follows from Eq.~\eqref{DeltaSigma}, the strain fluctuations become much larger than the average in the linear-response regime $\sigma \ll \sigma_L$:
    \be
   \Delta \sigma \gg  \langle \sigma \rangle=\sigma \quad \text{for} \quad \sigma \ll \sigma_L.
   \ee
This means that in order to obtain correct values of thermodynamic averages in this regime out of numerical simulations, one should exert a particular care to perform averaging over a sufficiently large statistical ensemble.

\section{Crossover to non-universal behavior with decreasing system size.}

Up to now we focused on the regime of large system sizes, $L \gg L_*$.
Let us now analyze a crossover from this regime to that of relatively small systems, $L \lesssim L_*$.
Expressing the anomalous deformation in the form
\be
\frac{Y_0\delta K_\beta}{2\sigma}= F_\beta (\sigma),
\label{FF}
\ee
we rewrite PR as follows [see Eq.~\eqref{nu-absolute}]:
\be
\nu=\frac{\nu_0-F_y(\sigma)}{1+F_x(\sigma)}.
\label{nuF}
\ee
We consider first the behavior of $\nu=\nu_L(\sigma)$  as a function of $L$
 exactly at $\sigma=0$.  Scaling properties  of $F_\beta$  can be understood within the approximation (\ref{G-SCSA-1}):
 \be
F_\beta (\sigma)=\frac{d_cT Y_0}{2 L^2} \sum\limits_{\mathbf q}
\frac{q_\beta^2 q_x^2}{\varkappa_q q^4 (\varkappa_q  q^4+\sigma q_x^2)}.
\label{F-sigma}
\ee
Evaluating the sum over momenta, we find
 \be
 F_\beta(0)  \sim
    \left\{ \begin{array}{ll}
        (L/L_*)^{2-2\eta},\ \ & L\gg L_*;
                                          \\
        (L/L_*)^{2}  , \ \ & L \ll L_*,
                                       \end{array}\right.
\ee
which yields the following result for the PR:
\begin{align}
\label{finite-size-corr1}
   & \nu_L(0) \approx
   \\
   \nonumber
    &\left\{ \begin{array}{lcl}
        \nu_\infty(0)   +  A_3\left({L_*}/{L}\right)^{2-2\eta}, \ \ & L\gg L_*;
   \\
    \nu_0- A_4 \left({L}/{L_*}\right)^{2},\ \ &L \ll L_*.
                                       \end{array}\right.
    \end{align}
Here $A_3\sim 1$ and $A_4 \sim 1$ are positive numerical coefficients, while $\nu_\infty(0)$ is
 the PR for $\sigma=0$ and $L=\infty$, with the limit $\sigma=0$  taken first.

Equation~\eqref{finite-size-corr1} is  general and valid for arbitrary dimensionality $d_c$ (i.e., arbitrary $\eta$). It is in agreement with  Ref.~\cite{nelson15}.
We see that with decreasing $L$, the PR increases from a negative value  $\nu_\infty(0)$ to the positive value
$\nu_0$ prescribed by the conventional elasticity theory.   The value  of $\nu_\infty(0)$ depends on $\eta$ and on BC.  For $\eta \to 0$, it is given by Eqs.~\eqref{per}, \eqref{free} and \eqref{zero} for three types of BC.   Equation~\eqref{finite-size-corr1} holds also for a disordered membrane, with a replacement  $\eta \to \eta_{\rm dis} \simeq \eta/4$,
see Sec.~\ref{sec:disordered} below.

\begin{figure}[ht]
\centerline{\includegraphics[width=0.9\linewidth]{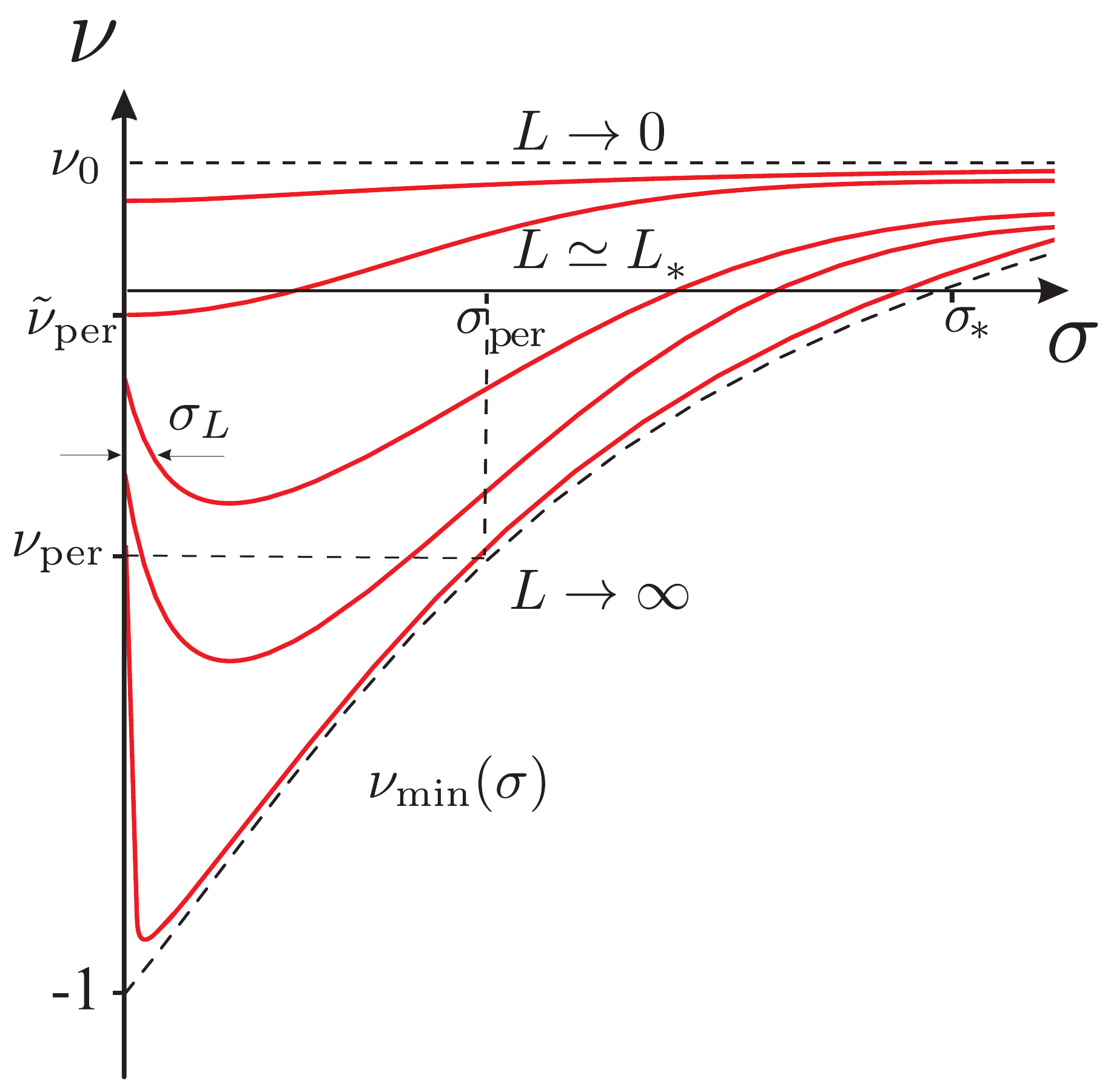}}
\caption{Evolution of the stress dependence of  absolute PR $\nu_L(\sigma)$ with system size $L$ for periodic boundary conditions. With increasing ratio $L/L_*$ the curve $\nu(\sigma)$ moves downwards. Lower dashed line: $L / L_* \to \infty$; upper dashed line: $L / L_* \to 0$.  The value $\sigma_{\rm per}$ is defined by condition $\nu_{\rm min} (\sigma_{\rm per}) = \nu_{\rm per}. $
For $\eta \to 0$, the limiting value $\nu_{\rm min}(0)$ is equal to $-1$, as shown in the plot. The values
of $\nu_{\rm per}$ and $\tilde \nu_{\rm per}$ are then given by Eqs.~\eqref{per} and \eqref{tilde-per}, respectively. For a generic case (including the physical case of $d_c=1$ with $\eta \approx 0.7 - 0.8 $) the numerical values are different but the qualitative behavior is expected to be the same.
}
\label{Fig4}
\end{figure}

\begin{figure}[ht]
\centerline{\includegraphics[width=0.9\linewidth]{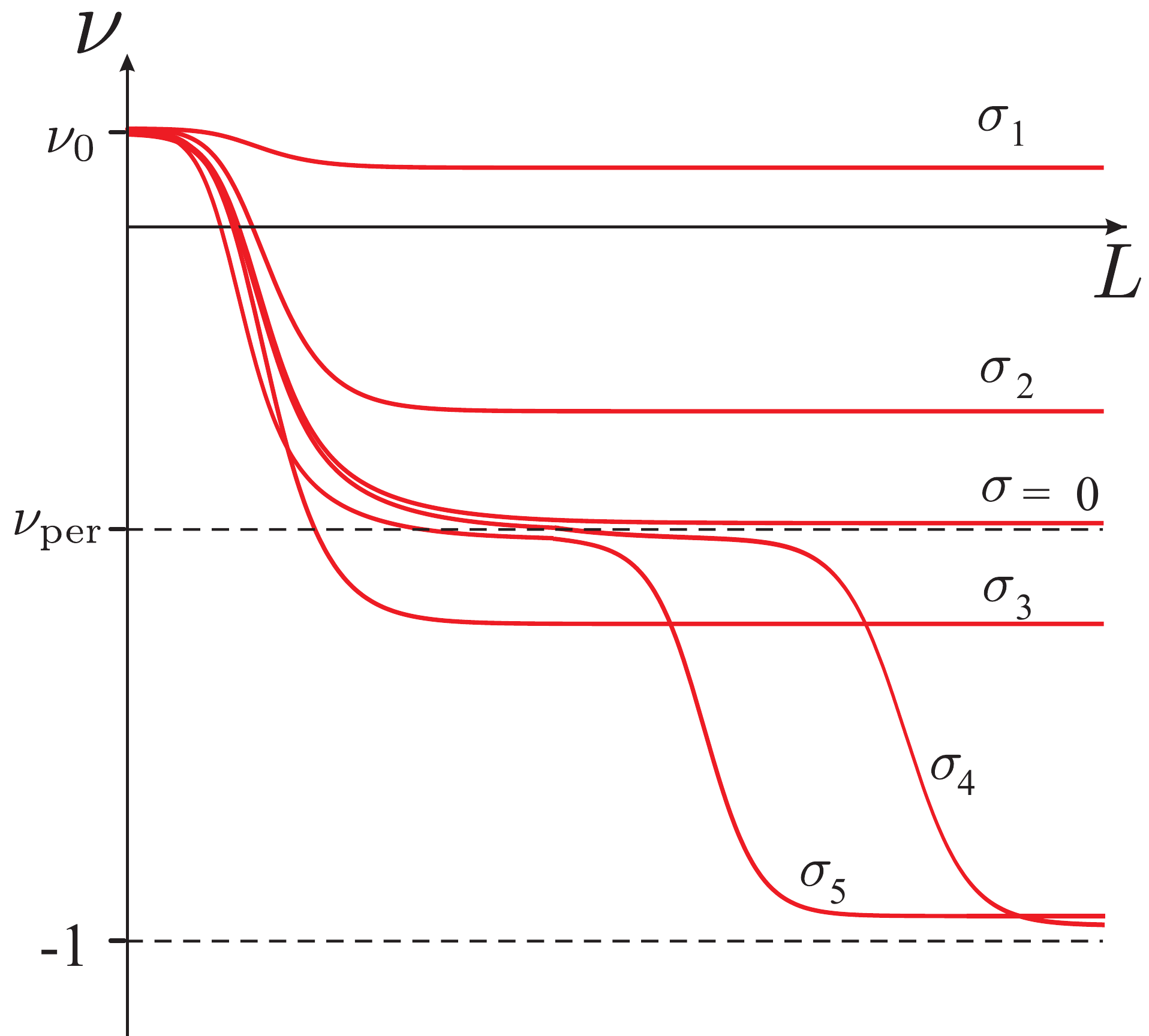}}
\caption{Length dependence of the absolute Poisson ratio  for  periodic BC and
different values of the applied  stress. The curves correspond to:
large stress,  $\sigma_1 \gg \sigma_*$,  intermediate stress, $\sigma_2 \gtrsim \sigma_{\rm per} \sim\sigma_*$ (with $\sigma_{\rm per}$ as defined in  Fig.~\ref{Fig4})),  relatively small stress, $\sigma_3\lesssim \sigma_{\rm per}$,   low stresses, $\sigma_4$ and $\sigma_5$ (with $\sigma_4<\sigma_5$), and  zero stress, $\sigma =0$.
For large stress $\sigma_1$,
the PR is close to $\nu_0$ within the whole interval of $L$.
The  curves, corresponding to  $\sigma \ll \sigma_*$ ($\sigma=\sigma_4$ and $\sigma=\sigma_5$)  show a well developed intermediate plateau at $\nu = \nu_{\rm per}$ and eventually saturate at $\nu =  \nu_{\rm min}(\sigma)$ which    approaches $\nu_{\rm min}(0)$.
For $\eta \to 0$, the value $\nu_{\rm min}(0)$ is equal to $-1$, as shown in the plot. The value of $\nu_{\rm per}$ is then given by Eqs.~\eqref{per}. For a generic case, the numerical values are different but the qualitative behavior is expected to be the same.
     }
\label{Fig5}
\end{figure}

Let  us now analyze what happens at small but finite stress,  $\sigma \ll \sigma_L$, assuming an arbitrary relation between $L$ and $L_*$.  This can be done in a controllable way  in the limit $\eta \to 0$. In this limit, expression \eqref{F-sigma}  becomes exact. As we demonstrated
in the previous section, the behaviour of the PR at small stress depends on BC.
For simplicity, we restrict ourselves to the  analysis  of the absolute PR of a membrane with periodic BC.
 Then  $$-\nu_\infty(0)=F_y(0)/F_x(0) = -\nu_{\rm per} $$  [see Eq.~\eqref{per}].
 Expanding $F_x$ and $F_y$  over $\sigma$ up to the first order and substituting these expansions into denominator and numerator of Eq.~\eqref{nuF}, we obtain
   \be
   \nu \approx \frac{\nu_0 + \nu_{\rm per}F_0 +\tilde \nu_{\rm per} \sigma  F_1}{1+F_0 +\sigma F_1}.
   \label{nu-FF}
   \ee
Here $F_0=F_x(0),$ $F_1=(dF_x/d\sigma)_{\sigma=0} $ and
\begin{align}
& \tilde \nu_{\rm per}=-\left(\frac{dF_y/d\sigma}{dF_x/d\sigma}\right)_{\sigma=0}= \nonumber
\\
& =   \frac{ \displaystyle 2 \sum \limits_{n=1}^{n=\infty} \sum \limits_{m=1}^{m=\infty}\frac{m^2n^4}{(n^2+m^2)^6}  }
{\displaystyle 2 \sum \limits_{n=1}^{n=\infty} \sum \limits_{m=1}^{m=\infty}\frac{ n^6}{(n^2+m^2)^6} + \sum \limits_{n=1}^{n=\infty} \frac{1}{n^6} } \simeq -0.03.
\label{tilde-per}
\end{align}
The sum in the last equation is obtained exactly in full analogy with Eq.~\eqref{per}.
It worth mentioning that the absolute  value of $\tilde  \nu_{\rm per}$ is numerically an order of magnitude smaller than a naive result,  $$\tilde \nu_{\rm per}^{\rm continous }  =  -\frac{ \langle n_x^4n_y^2\rangle_\varphi}{\langle n_x^6\rangle_\varphi}=-\frac15, $$
that one would get by using a continuous approximation, with all sums in Eq.~\eqref{tilde-per} replaced with the integrals.
  Equation \eqref{nu-FF} implies that at a certain value of $L/L_*$ the derivative $ (\p \nu/\p \sigma)_{\sigma \to 0}$ changes sign.   Differentiating Eq.~\eqref{nu-FF}  over $\sigma$, we find that the value of the PR at this point is  given exactly by  $\tilde \nu_{\rm per}$:
\be
 \left(\frac{d\nu}{d\sigma}\right)_{\sigma=0} =0 \quad \text{ for} \quad  \nu= \tilde \nu_{\rm per}.
\ee
Thus, for periodic BC, the dependence $\nu(\sigma)$ evolves from a non-monotonous to a monotonous one with
lowering $L/L_*$, as illustrated in Fig.~\ref{Fig4}. In Fig.~\ref{Fig5} we show how the dependence of $\nu$ on the system size $L$ evolves with increasing stress $\sigma$.

\section{Disordered membrane }
\label{sec:disordered}

In this Section, we discuss briefly a generalization of the results of this paper on the disordered case. A more detailed analysis will be presented elsewhere. Physically, the clean and disordered cases are quite similar. As was recently demonstrated  \cite{my-crump}, the bending rigidity of a strongly disordered membrane scales  in a power-law way in a wide interval of $q$:    $\varkappa_q \propto 1/q^{\eta_{\rm dis}}$. For small $d_c$, the critical index of the disordered problem is related to that of a clean system via  $\eta_{\rm dis} \simeq \eta/4$.
The power-law dependence of the effective isotropic stiffness,   \eqref{Hooke},  is also valid for strongly disordered membrane with the critical index $\alpha^{\rm dis}= \eta_{\rm dis}/(2-\eta_{\rm dis} ) \simeq \eta/(8-\eta)$.
Hence,  the clean and strongly disordered systems belong to different universality classes, i.e.,
exhibit power-law scaling of $\kappa_q$ characterized by different exponents, and  are thus characterized by distinct values of $\nu$.

\begin{figure}[h!]
\centerline{\includegraphics[width=0.85\linewidth]{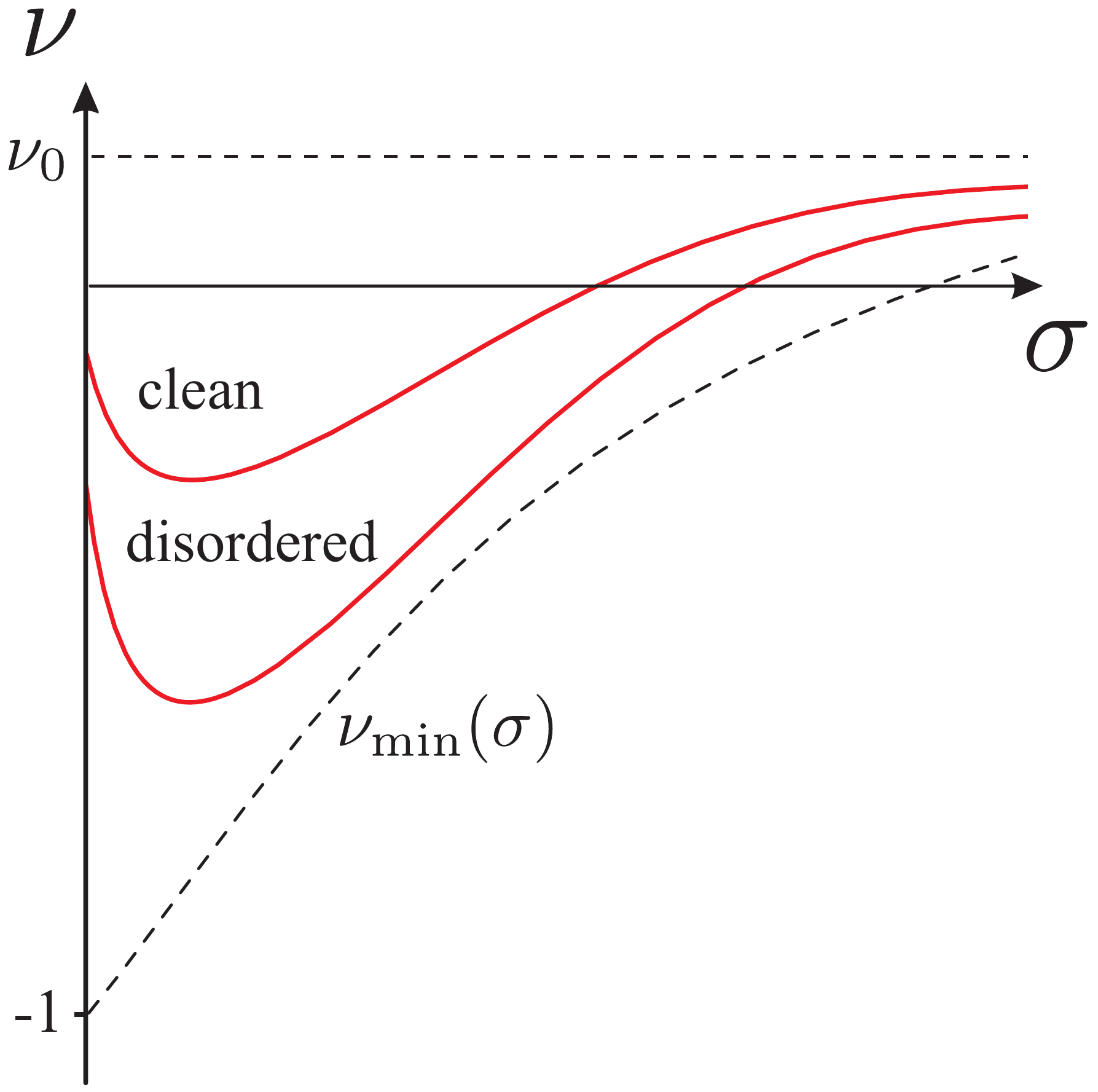}}
\caption{ Schematic plot of the stress dependence  of   PR of  a strongly disordered membrane as compared to a clean membrane with the same value of $\eta.$}
\label{Fig-dis}
\end{figure}

For a given value of $\eta$ (determined by the spatial dimension $d_c$) and the same BC,  we expect
the PR of a large ($L\gg L_*$) disordered  membrane to be lower than that for a clean membrane:
$\nu_{\rm dis}(\sigma) <\nu_{\rm clean} (\sigma) $, see Fig.~\ref{Fig-dis}.
Indeed, as has been demonstrated above, the PR tends to universal curve $\nu_{\rm min}$ with decreasing $\eta$.  Since  the effective $\eta$ for a disordered membrane is smaller than for a clean one, $\eta_{\rm dis} \simeq \eta/4 < \eta$,
the value of PR should be closer to this universal curve. This conclusion is in agreement with the numerical simulations that predicted a stronger auxetic behavior (i.e., more negative values of PR) for artificially disordered membranes \cite{Grima2015,Qin2017,Wan2017}, see Sec.~\ref{sec:intro}.

An important hallmark of the disordered case is mesoscopic fluctuations of the observables, in particular, of PR. These fluctuations become particularly prominent in the low-stress regime, $\sigma \ll \sigma_L $, when all the infrared divergencies are regularized by the system size, so that no self-averaging occurs. In particular, the mesoscopic fluctuations of PR  at $\sigma \ll \sigma_L $ should be of order unity.

\section{Summary and discussion}
\label{s6}

\begin{figure}[ht!]
\centerline{\includegraphics[width=0.9\linewidth]{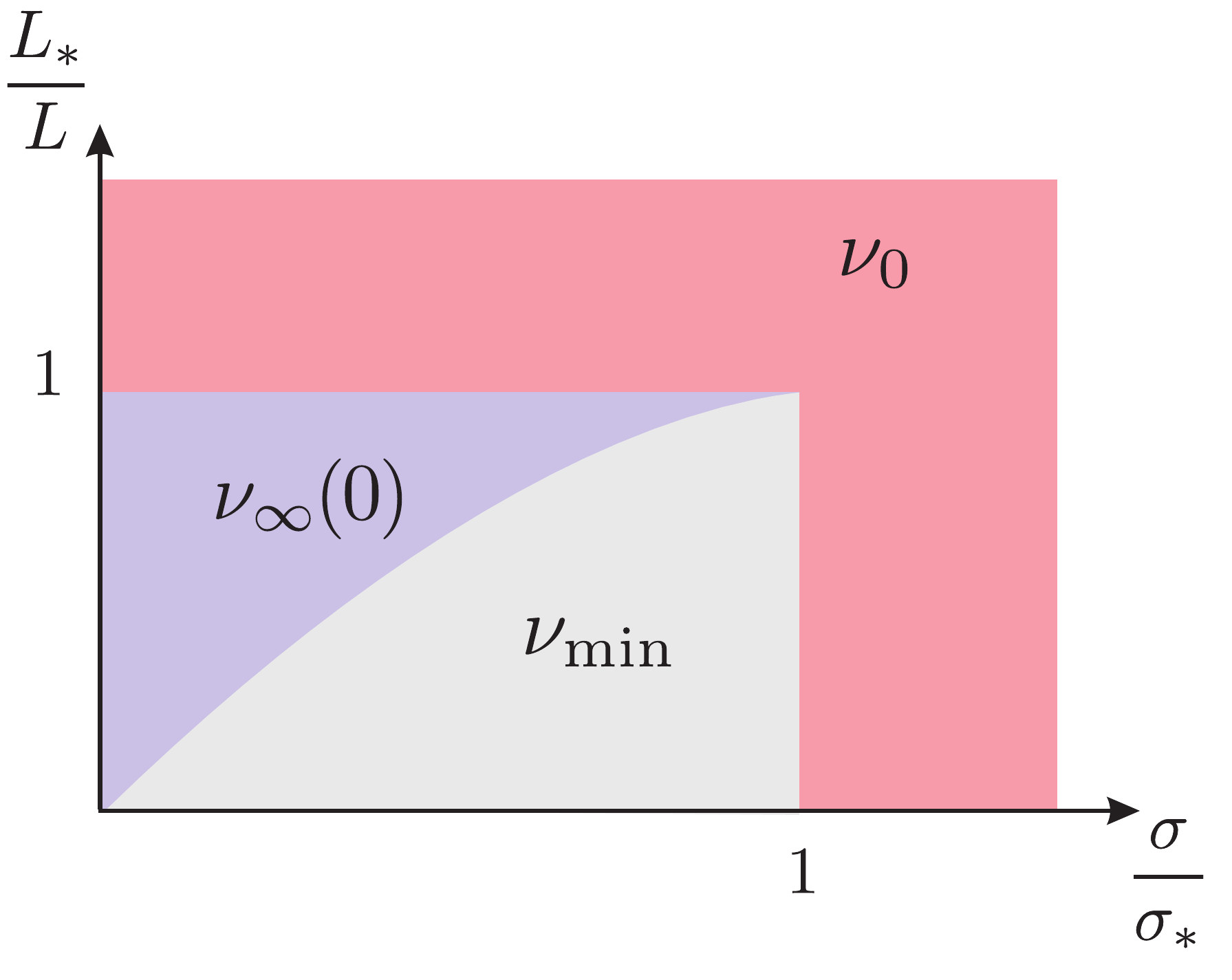}}
\caption{Regimes of asymptotic behavior of  Poisson ratio  in the parameter plane spanned by the stress and the system size. Crossovers between respective regimes take place around the lines $\sigma/\sigma_* \sim 1$, $L_*/L \sim 1$, and $\sigma/\sigma_* \sim (L_*/L)^{2-\eta}$.  }
\label{Fig3}
\end{figure}

To conclude, we have studied the system-size and stress dependence of the Poisson ratio of graphene (or, more generally, of a 2D membrane).  The ``phase diagram'' of various asymptotic regimes of the behavior of PR is presented in Fig.~\ref{Fig3}.
Our analysis, including the phase diagram, scaling, universality, as well as importance to distinguish between the absolute and differential PR, is valid for any spatial dimensionality $d_c$, including the physical case $d_c=1$. It has been supplemented by an evaluation of the PR in the limit of high dimensionality of the embedding space, $d_c \gg 1$, i.e., $\eta \ll 1$.

Our predictions  for a large system, $L \gg L_*$,  are schematically  summarized in  Fig.~\ref{Fig0}.   The red and blue curves represent the stress dependence of the absolute PR, $\nu$, and the differential PR, $\nu^{\rm diff}$, respectively.
One can observe the two characteristic scales of the stress, $\sigma_*$ and $\sigma_L$, which subdivide the $\sigma$ axis into three distinct regimes.

For  high stress, $ \sigma \gg  \sigma_*$,   non-linear effect are suppressed, the membrane obeys the conventional  linear Hooke's law, and the PR (both absolute and differential) is given by  its bare (material-dependent) value $\nu_0$,
\be
 \nu_0 = \frac{\lambda_0}{2\mu_0+\lambda_0}.
 \ee
 For graphene $\nu_0 \simeq  0.1$.

For low tensions, $\sigma \ll  \sigma_L$, elastic properties of the membrane are dominated by finite-size effects.
In this case, the membrane shows linear response with respect to external forces, so that the absolute and differential PR coincide: $\nu=\nu^{\rm diff}=\nu_\infty (0) $.  Here $\nu_\infty (0) $ stands for the following order of limits:  {\it first} one sends  $\sigma$ to zero and {\it next}   $L\to \infty$ (in fact, $L \gg L_*$ is sufficient).
An important prediction is a strong dependence of  $\nu_\infty (0)$ on boundary conditions.  For three types of BC---periodic, free and zero---we find that $\nu_\infty (0)$ is given, in the limit $\eta \to 0$,  by  $\nu_{\rm per}=-0.135,$  $\nu_{\rm free}=-0.075,$ and  $\nu_{\rm zero}=-0.735$, respectively. In this regime, the value of $\nu$ is universal in the sense that it does not depend on material parameters. It depends, however, on spatial dimensionality $d_c$ (i.e., on $\eta$) and on BC. These results for the low-tension regime are qualitatively consistent with numerical simulations of membranes in Refs. \cite{Zhang96,Bowick97,Bowick2001} which yielded negative values of PR and indicated importance of boundary conditions.

In the intermediate interval  $\sigma_L \ll \sigma \ll  \sigma_*$ the membrane falls into the universal non-linear regime, where the difference between the absolute and differential PR is essential, $\nu \neq \nu^{\rm diff}$.
For $\eta \to 0,$ both absolute and differential PR can be calculated analytically in a controllable way:
\be
\left\{  \begin{array}{c}
           \nu \to -1+ (1 + C_\Sigma)\eta +O(\eta^2),
            \\[0.3cm]
           \nu^{\rm diff}  \to -1/3 +0.008~ \eta +O(\eta^2),
         \end{array}
  \right.
  \ee
where $C_\Sigma$ is a numerical coefficient.
It is worth stressing again that the absolute and differential PRs in this regime have a high degree of universality: they depend only on the spatial dimensionality $d_c$ (or, equivalently, on $\eta$). We also notice that the numerical
coefficient in the first-order expansion of $\nu_{\rm diff}$ with respect to $\eta$ is very small, so that in the  physical case, $d_{\rm c}=1$, the differential PR may be expected to be relatively close to $-1/3$.  However, we cannot exclude at this stage a possibility that higher-order corrections are numerically larger. Therefore, high-precision numerical simulations would be highly desirable to check the above expectation.  

An interesting consequence of these findings is a non-monotonous dependence of the  PR on stress. As seen from Fig.~\ref{Fig0}, the absolute PR at small $\eta$ is  non-monotonous for any BC, because    $\nu_\infty (0)>-1$ for all BC. On the other hand, the differential PR is non-monotonous  for periodic and free BC (since $\nu_{\rm per}> -1/3$ and  $\nu_{\rm free} >-1/3$),  and monotonous for zero BC (since $\nu_{\rm zero}<- 1/3$).

We have further discussed the evolution of the above results with decreasing system size, when the system evolves towards the non-universal regime, $L < L_*.$   While the general tendency is  quite simple---both $\nu$ and $\nu^{\rm diff}$ tend to the non-universal value $\nu_0$, dependencies of PR on $\sigma$ for different $L$ and on $L$ for different $\sigma$ show interesting features, as illustrated in Figs.~\ref{Fig4} and \ref{Fig5}.
In particular,  the dependence of $\nu$ on $L$ for a  fixed stress demonstrates a wide plateau for sufficiently
low $\sigma$.

\begin{figure}[ht!]
\centerline{\includegraphics[width=1.0\linewidth]{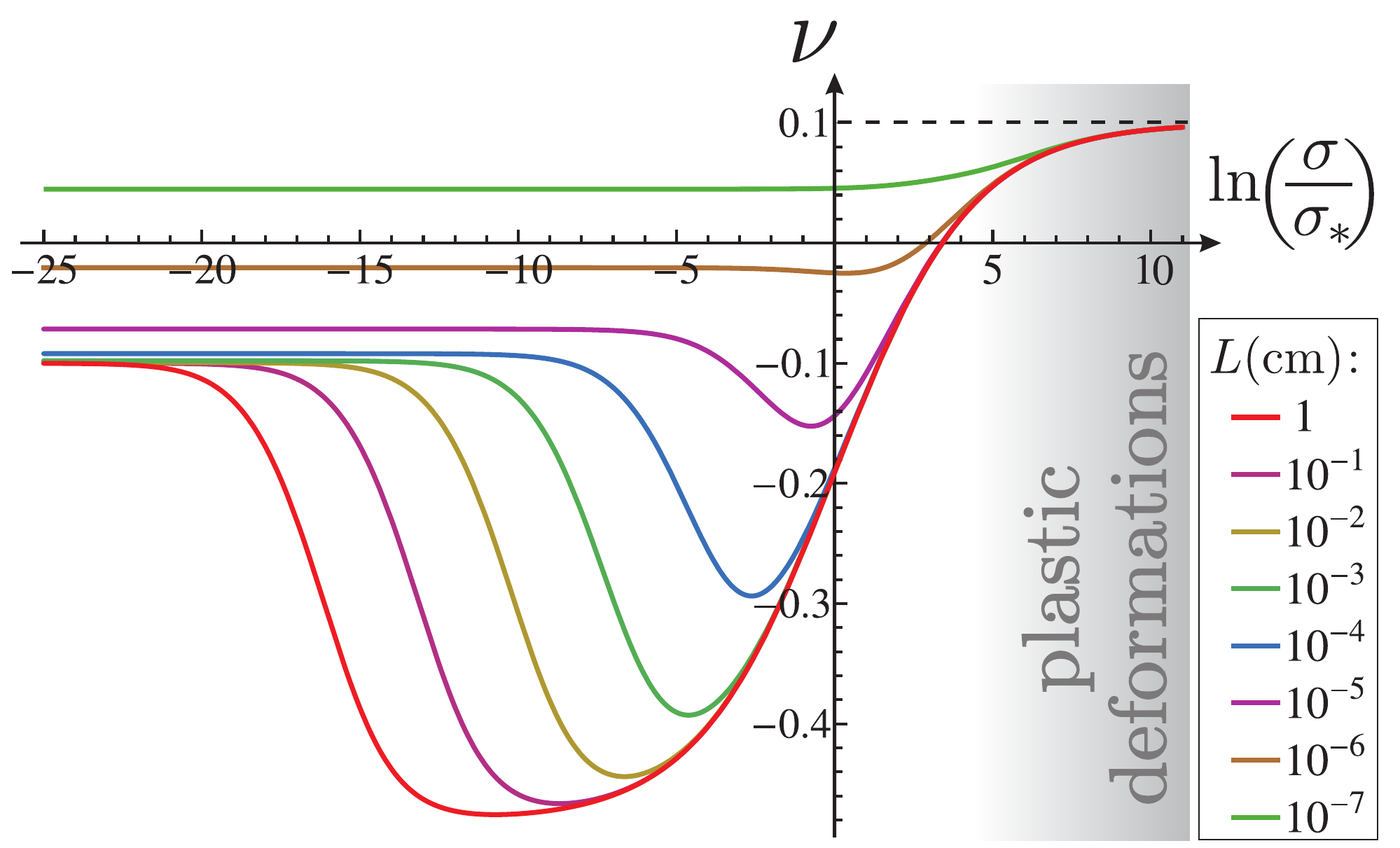}}
\caption{Stress dependence of the absolute Poisson ratio of free-standing graphene  at  $ T = 300 $ K in samples of different size (see text for details). Region marked by grey color corresponds to very large strains, $\sigma > 50\:$N/m, where a physical graphene membrane is affected by plastic deformations not included in our theory. }
\label{Fig-graphene}
\end{figure}

The results described above are applicable to generic membranes, including free-standing graphene. Qualitatively, different regimes  of behavior of the PR (schematically presented in Fig.~\ref{Fig0}) are better visualized if one uses the logarithmic scale for stress. In Fig.~\ref{Fig-graphene}, we present in this way the results for the absolute PR using the graphene parameters. Estimates have been  made for the room temperature in a very wide range of sample  sizes: from 10 \AA ~  to 1 cm. We used cyclic boundary conditions. The value of $\sigma_ *$ was estimated as $\sigma_ * =0.1$ N/m \cite{my-hooke}  and the  Ginzburg length as $L_* = 50 $ \AA \cite{my-crump}. The curves have been  obtained by interpolating between scaling dependencies that describe different regions.  The region of large stress corresponding  to plastic deformations (not included in our theory) is marked by grey color. The border between elastic and plastic deformations, $\sigma_{\rm plast}$ is estimated  as follows.  Plasticity comes into play   at deformations on the order of  20\% \cite{plasticity}.   Since the Young modulus of graphene equals 340 N/m,   we estimate $\sigma_{\rm plast} \simeq 50\:$ N/m, thus obtaining $\ln(\sigma_{\rm{plast}}/\sigma_*) \simeq 6.$   It is seen from Fig.~\ref{Fig-graphene} that in order to observe all three regimes (``mesoscopic'', universal, and conventional elasticity) one has  to pass a rather wide interval of stresses.  It is also worth noting that the region of conventional elasticity has a strong overlap with the plasticity region. In other words, the elasticity of graphene is mainly anomalous.   

We  have also discussed qualitatively the case of a disordered membrane. The  key difference is that the   effective $\eta$ for  a disordered membrane is smaller than for clean one, $\nu_{\rm dis} \simeq \nu/4<\eta.$     Since the dependence $\nu $  on $\sigma$  tends to $\nu_{\rm min} (\sigma)$ with decreasing $\eta$ (for $L\to \infty$), the value of the PR for a disordered membrane  should be closer to the universal curve $\nu_{\rm} (\sigma)$, see Fig.~\ref{Fig-dis}.
A hallmark of a disordered membrane is strong mesoscopic (sample-to-sample) fluctuations of PR in the linear-response regime, $\sigma \ll \sigma_L$. Our conclusions are in agreement with  numerical simulations which found a more pronounced auxetic behavior for artificially disordered graphene membranes \cite{Grima2015,Qin2017,Wan2017}. Furthermore, Fig.~3 of Ref.~\cite{Grima2015} shows a strongly non-monotonous behavior of PR as a function of applied stress for disordered graphene, in full consistency with our Fig.~\ref{Fig0}. On the other hand, for the case of clean graphene,  an initial reduction of PR for smallest stresses was observed  in  Ref.~\cite{Grima2015} (see upper panel of their Fig.~3).  The reason for this was likely an insufficient system size, see Fig.~\ref{Fig4} of our work. The disorder reduces the Ginzburg length $L_*$, thus allowing one to probe better the universal regime $L \gg L_*$ for a given system size $L$. 

Our work paves the way for detailed studies of PR and related properties of graphene and other 2D membranes. Since the physical situation of $d_c=1$
($\eta \approx 0.7-0.8$) does not belong to the regime $d_c \to \infty$ ($\eta\to 0$) where our analytical calculations
of the asymptotic values of PR are fully controllable, systematic numerical simulations would be of great interest. Identification of the three regimes (see Figs.~\ref{Fig0} and \ref{Fig3}) in such simulations is expected to be a feasible, although rather challenging task. A more ambitious goal is a sufficiently precise determination of the asymptotic values of the absolute and differential PR in the non-linear universal regime, as well as of PR in the linear-response regime for various BC.

We hope that our work will also stimulate further experimental activities on nanomechanics of clean and disordered graphene (and related 2D materials), both in the  linear and non-linear regime with respect to the applied stress.  In particular, an experimental realization  of regimes of auxetic behavior identified and analyzed in our work would be of great interest.

\section{Acknowledgments}

We thank   K.I.  Bolotin  for useful comments. The work of  I.S.B., I.V.G, V.Yu.K, and A.D.M was supported by the Russian Science Foundation (grant No. 14-42-00044).  J.H.L. and M.I.K. acknowledge funding from the European Unions Horizon 2020 research and innovation programme 206 under grant agreement No. 696656 GrapheneCore1.

\appendix
\section{Derivation of the free energy  and balance equations of anisotropically loaded membrane}
\label{App-balance}

In this Appendix, we present a derivation of Eqs.~\eqref{tensor-hooke} and \eqref{M} of the main text as well their generalization to the case of a membrane of arbitrary dimensionality $D$.

We write the  energy functional \eqref{E} in terms of the components of the stretching vector $\xi_x,\xi_y$ [see Eq.~\eqref{deformations}], in-plane fluctuations $\mathbf u,$  and out-of-plane fluctuations $\mathbf h$:
\be
E=L^2 \sum \limits_{\alpha\beta}\left[ \frac{\mu_0}{4} \delta_{\alpha\beta} +\frac{\lambda_0}{8}\right]
\left[s_\alpha s_\beta  - K_\alpha^0 K_\beta^0 \right] + E_0 (\mathbf u,\mathbf h),
\label{energy-functional}
\ee
where
\be
K_\alpha^0=\overline{ (\p_\alpha \mathbf h)^2} =\int \frac{d^2 x}{L^2}(\p_\alpha \mathbf h)^2,
\label{K-L}
\ee
[here $ \overline{(\cdots)}$  stands for the spatial averaging],
\be
s_\alpha=\xi_\alpha^2-1+K_\alpha^0,
\ee
and
\be
E_0 (\mathbf u,\mathbf h)= \int d^2x\left\{  \frac{\varkappa}{2} (\Delta\mathbf h )^2+\mu_0 u_{ij}^2+\frac{\lambda_0}{2}u_{ii}^2   \right\}.
\label{Fdis}
\ee
Here
$${u_{\alpha\beta}=\left(\xi_\beta \p_\alpha u_\beta+\xi_\alpha \p_\beta u_\alpha +\p_{\alpha}{\mathbf h}\p_\beta{\mathbf h}\right)/2}
 $$
is the strain tensor (the rule of summation over repeated indices does not apply).
For small deformations, one can put $\xi_x =\xi_y=1$ in the strain tensor, so that
$E_0 (\mathbf u,\mathbf h) $ coincides with the conventional expression for elastic energy of nearly flat membrane.

The next step is to calculate free energy corresponding to energy functional Eq.~\eqref{energy-functional}:
  \be
  F=-T \ln \left( \int \{ d\mathbf u d\mathbf h\}e^{-E/T} \right).
  \ee
This can be done in an analogy with isotropic case (see  technical details  in the Supplementary Materials of Ref.~\cite{my-hooke}).
We
decouple the term $s_\alpha s_\beta$ with the use of the auxiliary fields $\chi_\alpha$:
\BEA
&&\exp \left[ -\frac{L^2}{4T} \left( \mu_0\delta_{\alpha\beta}  +\lambda_0/2\right)s_\alpha s_\beta\right]
\label{ax}
\\
\label{ax1}
&&\!\propto\!\!\int \!d\chi_\alpha \exp\left\{\!\frac{L^2}{2T}\sum\limits _{\alpha\beta} \left[   \delta_{\alpha\beta}\left(i s_\alpha \chi_\beta -\frac{\chi_\alpha\chi_\beta}{2\mu}\right)\! \right. \right.
\\
&&
\left. \left.
+\! \frac{\lambda_0 ~\chi_\alpha \chi_\beta}{4\mu_0(\mu_0+\lambda_0)} \right]\!\right\}.
\nonumber
\EEA
Evaluating the integrals over the in-plane modes $\bf u$ and out-of plane modes $\bf h$, we find
\BEA
&&\Phi(\sigma,\xi)=\frac{L^2}{2}\left\{ \sum\limits_\alpha \left[ \sigma_\alpha (\xi_\alpha^2 -1) -\frac{\sigma_\alpha^2}{2\mu_0}\right]\right.
\nonumber
\\
&&\left.+\frac{\lambda_0}{4\mu_0(\mu_0+\lambda_0)} \left(\sum\limits_\alpha \sigma_\alpha\right)^2\right\}
\nonumber
\\
&&+\frac{T d_c}{2}\sum\limits_\mathbf q \ln\left(\varkappa_q q^4 + \sum\limits_\alpha \sigma_\alpha q_\alpha^2\right),
\label{Phi-xi}
\EEA
where $\sigma_\alpha = - i\chi_\alpha^{(0)}$ and $\chi_\alpha^{(0)}$ correspond to stationary phase condition for integral over $d\chi_\alpha$.  This  condition, $\partial \Phi(\sigma,\xi) /\partial \sigma_\alpha=0,$
yields the balance equations for a 2D membrane as presented in Eqs.~\eqref{tensor-hooke} and \eqref{M} of the main text.
Inserting the corresponding equilibrium values $\sigma_\alpha(\xi_x,\xi_y)$ in the functional
  $\Phi(\sigma_x, \sigma_y,\xi_x,\xi_y)$, we find the free energy $F(\xi_x,\xi_y)$ as a function of the stretching vector $(\xi_x,\xi_y)$.
The stationary-point value $\sigma_\alpha(\xi_x,\xi_y)$ determines the physical stress. Indeed, it is easy to check that
$
\sigma_{\alpha}= ({1}/{L^2\xi_\alpha})
{\partial {F}}/{\partial \xi_\alpha} \quad \text{(no summation over $\alpha$)},
$
i.e., $\sigma$ is conjugate to the strain $\xi$. Deep in the flat phase, $\xi_\alpha \approx 1$ and we find
\begin{equation}
\sigma_{\alpha}=\frac{1}{L^2}
\frac{\partial {F}}{\partial \xi_\alpha}.
\label{eq:stress}
\end{equation}

These results can be straightforwardly generalized to the case of a membrane of an arbitrary dimensionality $D$.
We assume that the membrane is loaded in a certain direction by uniaxial stress, with equal deformations in other $D-1$ directions.  First, we neglect anomalous deformations.  We get then
the following matrix of elastic constants  of a membrane:
\be
\hat M =\left(
          \begin{array}{cc}
            2\mu_0+\lambda_0 & \lambda_0 \\
            \lambda_0 (D-1) & 2\mu_0+\lambda_0 (D-1), \\
          \end{array}
        \right)
\label{M1}
\ee
which generalizes Eq.~\eqref{M} of the main text.
Using a standard definition of elastic moduli, we thus obtain
 \BEA
 &&
 B=\lambda_0 +\frac{2\mu_0}{D},
 \\
 &&
 Y_0=\frac{2\mu_0 (2\mu_0 +D \lambda_0)}{ 2\mu_0 +\lambda_0(D-1) },
 \\
 &&
 C_{11}=2\mu_0+\lambda_0,
 \\
 && \nu_0=\frac{\lambda_0}{2\mu_0 +\lambda_0 (D-1)}.   \label{nu_0_D}
 \EEA
 These equations can be extended to include anomalous deformations.
 In particular,
 the differential PR $\nu^{\rm diff}$ is expressed in terms of the renormalized elastic parameters
 $\mu^{\rm diff}$ and $\lambda^{\rm diff}$ by a formula that has the same form as Eq.~\eqref{nu_0_D}.
 In the universal non-linear regime (where the  anomalous deformations dominate), and in the limit $d_c\to \infty$,
 the results of Ref. \cite{Doussal} apply, yielding the invariant manifold of the  Lam{\'e} coefficients,
 \be
 \lambda^{\rm diff}=-\frac{2 \mu^{\rm diff}}{ D+2},\quad d_c=\infty,
 \ee
 cf. Eq.~\eqref{lam-diff} for $\Pi_+=\Pi_-$.
  Hence,  we get
    \be
    \nu^{\rm diff}=-\frac{1}{3} \quad\text{for}\quad d_c = \infty.
    \ee
Thus, for $d_c\to \infty$, the value $-1/3$ of the differential PR in the non-linear universal regime is independent of the membrane dimensionality $D$. As explained in the main text for the case of 2D membranes, this value gets modified when one considers a finite dimensionality $d_c$.


\begin{thebibliography}{99}

\bibitem{evans} K. E. Evans, M. A. Nkansah, I. J. Hutchinson, and  S. C. Rogers,
Nature  {\bf 353}, 124 (1991).

\bibitem{pyrite}  A.E.H. Love, {\it A Treatise on the Mathematical Theory of Elasicity } (Dover, New York, 4th ed., p.163, 1944).

\bibitem{foam}
	R. Lakes,
Science
{\bf  235},  1038 (1987).

\bibitem{auxetic}
J.-W. Jiang, S. Y. Kim, and H. S. Park,  Applied  Physics Reviews {\bf 3}, 041101 (2016).

\bibitem{Geim}  K.S.\ Novoselov, A.K.\ Geim, S.V.\ Morozov, D.\ Jiang, Y.\ Zhang, S.V.\ Dubonos, I.V.\ Grigorieva, and A.A.\ Firsov,
Science \textbf{306}, 666 (2004).

\bibitem{Geim1} K.S.\ Novoselov, A.K.\ Geim, S.V.\ Morozov, D.\ Jiang, M.I.\
Katsnelson, I.V.\ Grigorieva, S.V.\ Dubonos, and A.A.\ Firsov, Nature
\textbf{438}, 197 (2005).

\bibitem{Kim} Y.\ Zhang, Y.-W.\ Tan, H.L.\ Stormer, and P.\ Kim, Nature
\textbf{438}, 201 (2005).

  \bibitem{geim07}  A.K.\ Geim and K.S.\ Novoselov, Nature Materials {\bf 6},
183 (2007).

\bibitem{graphene-review}  A.H.\ Castro Neto, F.\ Guinea, N.M.R.\ Peres,
K.S.\ Novo\-selov, and A.K.\ Geim, Rev. Mod. Phys.  {\bf 81}, 109 (2009).

\bibitem{review-DasSarma}
S. \ Das Sarma, S. \ Adam, E. H. \ Hwang, and E.\ Rossi,
Rev. Mod. Phys. {\bf 83}, 407 (2011).

\bibitem{review-Kotov}
V.N. \ Kotov, B. \ Uchoa, V.M. \ Pereira, F. \ Guinea, and A. H. \ Castro Neto,
Rev. Mod. Phys. {\bf 84}, 1067  (2012).

\bibitem{book-Katsnelson} M.I. Katsnelson,
{\it Graphene: Carbon in Two Dimensions} (Cambridge University Press, 2012).

\bibitem{review-Katsnelson}  M.I. Katsnelson and A. Fasolino,
Acc. Chem. Res.  {\bf 46},  97  (2013).

\bibitem{book-Wolf}
E. L. Wolf, {\it  Graphene:
A New Paradigm in Condensed Matter and Device Physics} (Oxford University Press, 2014).

\bibitem{book-Roche}
 L.E.F. Foa \ Torres,     S.\ Roche, and
    J.-C.\ Charlier,  {\it
Introduction to Graphene-Based Nanomaterials
From Electronic Structure to Quantum Transport}  (Cambridge  University Press, 2014).

 \bibitem{Nelson} D.\ Nelson, T.\ Piran, and S.\ Weinberg (Eds.) {\it Statistical Mechanics of Membranes and Surfaces} (World Scientific, Singapore, 1989)

\bibitem{Cao2014}
G. Cao,   Polymers  {\bf 6}, 2404 (2014).

\bibitem{Politano2015}
A. Politano and G. Chiarello, Nano Research, {\bf  8},  1847 (2015).

 \bibitem{Zakharchenko2009} K. V. Zakharchenko, M. I. Katsnelson, and A. Fasolino, Phys.
Rev. Lett.  {\bf 102}, 046808  (2009).

\bibitem{katsnelson16} J. H. Los, A. Fasolino, and M. I. Katsnelson,  Phys. Rev. Lett. {\bf 116}, 015901 (2016).

\bibitem{Li}  G. Gui, J. Li, and J. Zhong, Phys. Rev. B {\bf 78}, 075435  (2008).  

\bibitem{Grima2015}
J. N. Grima, S. Winczewski, L. Mizzi, M. C. Grech, R. Cauchi,
R. Gatt, D. Attard, K. W. Wojciechowski, and J. Rybicki,
Adv. Mater.  {\bf 27}, 1455 (2015).

\bibitem{Qin2017}
H. Qin,  Y. Sun,  J. Z. Liu,
M. Lia,  and  Y. Liu,
Nanoscale {\bf 9}, 4135  (2017).

\bibitem{Wan2017}
J. Wan, J.-W. Jiang, and H. S. Park,
 Nanoscale {\bf  9}, 4007  (2017).

\bibitem{Jiang2016}
J.-W. Jiang, T. Chang, X. Guo, and H. S. Park,
Nano Letters  {\bf 16}, 5286  (2016).

\bibitem{Ulissi2016}
Z. W. Ulissi, A. G. Rajan, and M. S. Strano,
ACS Nano {\bf  10}, 7542 (2016).


\bibitem{Park2016}
J.-W. Jiang, and H. S. Park,
Nano Letters  {\bf 16}, 2657 (2016).


\bibitem{Wu2015}
Y. Wu, N. Yi, L. Huang, T. Zhang, S. Fang, H. Chang, N. Li, J. Oh,
J. A. Lee,  M. Kozlov, A. C. Chipara, H. Terrones, P. Xiao, G. Long,
Y. Huang, F. Zhang, L. Zhang, X. Lepro, C. Haines, M. D. Lima, N. P. Lopez,
L. P. Rajukumar, A. L. Elias, S. Feng, S. J. Kim, N.T. Narayanan, P. M. Ajayan,
M. Terrones, A. Aliev, P. Chu, Z. Zhang, R. H. Baughman, and Y. Chen,
Nature Commun. {\bf 6},  6141 (2015).


\bibitem{Ho2016}
V. H. Ho, D. T. Ho, S.-Y. Kwon, and S. Y. Kim,
Phys. Status Solidi B  {\bf 253}, 1303 (2016).



\bibitem{Doussal} P.\ Le Doussal and L.\ Radzihovsky, Phys. Rev. Lett. 
{\bf 69}, 1209 (1992).

\bibitem{nelson13} A. Kosmrlj and D. R. Nelson, Phys. Rev. E {\bf 88}, 012136 (2013); 
Phys. Rev. E {\bf 89}, 022126 (2014).

\bibitem{SCSA-Review} P. Le Doussal and L. Radzihovsky, arXiv:1708.05723.

\bibitem{nelson15} A. Kosmrlj and D. R. Nelson, Phys. Rev. B  {\bf 93}, 125431 (2016).


 \bibitem{Zhang96} Z. Zhang,   H. T. Davis,  and D. M. Kroll,
Phys. Rev. E   {\bf  53}, 1422 (1996).


\bibitem{Bowick97} M. Falcioni,  M. J. Bowick, E. Guitter,  and G. Thorleifsson,
Europhys. Lett. {\bf 38}, 67 (1997).

\bibitem{Bowick2001} M. Bowick, A. Cacciuto, G.  Thorleifsson, and A. Travesset,
Phys. Rev. Lett. {\bf  87}, 148103 (2001).

\bibitem{footnote-nelson} Recently, numerical simulations of Ref.~\cite{nelson13} indicated a positive PR for warped 2D membranes at zero temperature, whereas the scaling exponents were found to be in agreement with their SCSA values. The authors of Ref.~\cite{nelson13} argued that the SCSA might be insufficient for calculating the PR.

\bibitem{mermin}
N. D. Mermin and H. Wagner, Phys. Rev. Lett. {\bf 17}, 1133 (1966);
N.D. Mermin,  Phys. Rev. {\bf 176}, 250 (1968).

\bibitem{landau} L.D.\ Landau and E.M.\ Lifshitz, {\it Statistical Physics, Part 1} (Pergamon Press, Oxford, 1980).


\bibitem{Nelson0} D.R.\ Nelson and L.\ Peliti, J. Phys. (Paris) {\bf 48}, 1085 (1987).

\bibitem{Crump1} Y.\ Kantor and D.R.\ Nelson, Phys.\ Rev.\ Lett.\ {\bf 58}, 2774 (1987); Phys.\ Rev.\ A\ {\bf 36}, 4020 (1987);

\bibitem{NelsonCrumpling} M.\ Paczuski, M.\ Kardar, and D.R.\ Nelson, Phys.\ Rev.\ Lett.\ {\bf 60}, 2638 (1988).

 \bibitem{david1} F.\ David and  E.\ Guitter, Europhys.  Lett. {\bf 5}, 709 (1988).

\bibitem{buck} E. \ Guitter, F.\ David, S.\ Leibler, and  L.\ Peliti,    Phys. Rev. Lett.
{\bf 61}, 2949 (1988).

 \bibitem{Aronovitz89} J.A.\ Aronovitz and T.C.\ Lubensky, Phys. Rev. Lett. {\bf 60}, 2634 (1988).

\bibitem{david2} E. \ Guitter, F.\ David, S. \ Leibler, and  L. \ Peliti,    J. Phys. France {\bf 50} 1787 (1989).

 \bibitem{lower-cr-D2} J. Aronovitz, L. Golubovi{\'c},  and  T.C. Lubensky,   J. Phys. France {\bf 50} 609 (1989).

 \bibitem{d-large} M. Paczuski  and M. Kardar,  Phys. Rev. A {\bf 39}, 6086 (1989).

\bibitem{disorders} L.\ Radzihovsky and D.R.\ Nelson, Phys. Rev. A {\bf
44}, 3525 (1991).

 \bibitem{disorder-imp} D.R.\ Nelson and L.\ Radzihovsky,
Europhys.  Lett. {\bf 16}, 79 (1991).

\bibitem{Gompper91} G.\ Gompper and D.M.\ Kroll, Europhys. Lett. {\bf 15},
783 (1991).

 \bibitem{RLD}  L.\ Radzihovsky and P.\ Le Doussal,  J.Phys.  I France {\bf 2} 599 (1992).

 \bibitem{disorders-Morse-Grest} D.C. Morse,  T.C. Lubensky, and  G.S. Grest, Phys. Rev. A {\bf 45}, R2151 (1992).

\bibitem{RLD1} P.\ Le Doussal and L.\ Radzihovsky, Phys. Rev. B {\bf 48}, 3548 (1993).

 \bibitem{Bowick96} M.J.\ Bowick, S.M.\ Catterall, M.\ Falcioni, G.\
Thorleifsson, and K.N.\ Anagnostopoulos, J. Phys. I France {\bf 6}, 1321 (1996).

\bibitem{Los-PRB-2009}
J.H.\ Los,  M.I.\ Katsnelson,  O.V.\ Yazyev,  K.V.\ Zakharchenko, and A.\ Fasolino,
 Phys. Rev. B  {\bf 80}, 121405(R) (2009).

 \bibitem{eta1} J.-P.\ Kownacki, and D.\ Mouhanna, Phys. Rev. E {\bf 79}, 040101(R) (2009).

\bibitem{Gazit1} D. \ Gazit, Phys. Rev. E {\bf 80}, 041117 (2009).

\bibitem{Gazit2} D. \ Gazit, Phys. Rev. B {\bf 80}, 161406(R) (2009).

\bibitem{Hasselmann} F.L.\ Braghin and N. \ Hasselmann, Phys. Rev. B {\bf 82}, 035407 (2010).


 \bibitem{kats1} V.V. \ Lebedev and E.I. \  Kats, Phys. Rev. B {\bf 85}, 045416 (2012).

 \bibitem{kats2} E.I. \ Kats and V.V. \ Lebedev,   Phys. Rev. B {\bf 89}, 125433 (2014).

\bibitem{Amorim} B. \ Amorim, R. \ Rold{\'a}n, E. \ Cappelluti, A. \ Fasolino, F. \ Guinea, and M. I. \
 Katsnelson, Phys. Rev. B {\bf 89}, 224307  (2014).

\bibitem{kats3} E. I.  \ Kats and V. V.  \ Lebedev, Phys. Rev. E  {\bf 91}, 032415 (2015).

\bibitem{lee} C. \ Lee, X. \ Wei, J.W.\ Kysar, and J. \ Hone, Science {\bf 321}, 385 (2008).
\bibitem{metten}  D. \ Metten, F. \ Federspiel, M. \ Romeo, and S. \ Berciaud, Phys. Rev. Applied {\bf 2}, 054008 (2014).

\bibitem{blees15}
M. K. Blees, A.W. Barnard, P. A. Rose, S. P. Roberts, K. L. McGill, P. Y. Huang, A. R. Ruyack, J. W. Kevek, B. Kobrin, D.A. Muller, and  P. L. McEuen, Nature {\bf 524}, 204 (2015).

\bibitem{lopez-polin15}
G. Lopez-Polin, C. Gomez-Navarro, V. Parente, F. Guinea, M. I. Katsnelson, F. Perez-Murano, and J. Gomez-Herrero, Nature Physics {\bf 11}, 26 (2015);
G. Lopez-Polin, M. Jaafar, F. Guinea, R. Roldan, C. Gomez-Navarro, and J. Gomez-Herrero,
Carbon {\bf 124}, 42, (2017).

\bibitem{nicholl15} R.J.T. Nicholl, H. J. Conley, N. V. Lavrik, I. Vlassiouk, Y. S. Puzyrev, V. P. Sreenivas, S. T. Pantelides, and K. I. Bolotin, Nature Commun. {\bf 6}, 8789 (2015).

\bibitem{my-cond} I. V. Gornyi, V. Yu. Kachorovskii, and A. D. Mirlin,
Phys. Rev. B  {\bf 86}, 165413 (2012).

\bibitem{Zhu}  T. Zhu and E. Ertekin, Phys. Rev. B {\bf 90}, 195209 (2014);  
Phys. Rev. B {\bf  91}, 205429 (2015).

\bibitem{my-crump} I.V. Gornyi, V. Yu. Kachorovskii,  and A. D. Mirlin,
Phys. Rev. B {\bf 92}, 155428 (2015).

\bibitem{my-hooke} I.V.\ Gornyi, V. Yu.\ Kachorovskii,  and A. D.\ Mirlin, 2D Materials {\bf 4},  011003  (2017).


\bibitem{my-quantum}  I. S. Burmistrov, I. V. Gornyi, V. Yu. Kachorovskii, M. I. Katsnelson, 
and A. D. Mirlin,
Phys. Rev. B {\bf 94}, 195430 (2016).

\bibitem{lifshitz} I. M. Lifshitz, Sov. Phys. JETP {\bf 52}, 472 (1952).

\bibitem{my-Ward}  I. S. Burmistrov, I. V. Gornyi, V. Yu. Kachorovskii,  and A. D. Mirlin,   arXiv:1801.05053.

\bibitem{plasticity} M. Topsakal and S. Ciraci,
Phys. Rev. B {\bf 81}, 024107 (2010).

\end{thebibliography}
 \end{document}